\documentclass[manuscript,screen,nonacm]{acmart}
\usepackage{epsfig}
\usepackage{graphicx}

\usepackage{amsmath}
\usepackage{tabularx}
\usepackage{lscape}
\usepackage{longtable}
\usepackage{wrapfig}
\usepackage{arydshln}
\usepackage{xcolor}

\usepackage{dirtytalk}

\newcommand{\subf}[2]{%
  {\small\begin{tabular}[t]{@{}c@{}}
   \mbox{}\\[-\ht\strutbox]
   #1\\#2
   \end{tabular}}%
}

\usepackage{rotating}
\usepackage{afterpage}
\usepackage{comment}
\usepackage{color,soul}
\usepackage{tikz}
\usetikzlibrary{positioning}

\usepackage{pgfplots}
\pgfplotsset{width=7cm, compat=1.9}
\usepackage{pgf-pie}

\newlength\Textht
\usetikzlibrary{trees}

\usepackage{subcaption}
\usepackage{xcolor}
\usepackage{threeparttable}
\usepackage{algorithm}
\usepackage{algorithmic}
\usepackage{enumitem}
\usepackage{multirow}
\usepackage{lipsum}
\usepackage{comment}

\usepackage{xspace}
\makeatletter
\DeclareRobustCommand\onedot{\futurelet\@let@token\@onedot}
\def\@onedot{\ifx\@let@token.\else.\null\fi\xspace}

\newcolumntype{P}[1]{>{\centering\arraybackslash}p{#1}}

\makeatletter
\renewcommand*{\p@section}{\S\,}
\renewcommand*{\p@subsection}{\S\,}
\makeatother

\newcommand{\erhao}[1]{\fontsize{11pt}{\baselineskip}\selectfont}

\usepackage{url}
\usepackage[utf8]{inputenc}
\usepackage{booktabs}
\usepackage[american]{babel}
\usepackage[T1]{fontenc}
\usepackage{balance}

\AtBeginDocument{%
  }

\setcopyright{acmcopyright}
\copyrightyear{2023}
\acmYear{2023}
\acmDOI{XXXXXXX.XXXXXXX}

\begin{document}

\title{Physics-Informed Computer Vision: A Review and Perspectives}

\author{Chayan Banerjee}
\email{c.banerjee@qut.edu.au}
\affiliation{%
  \institution{Queensland University of Technology}
  \streetaddress{2 George Street}
  \city{Brisbane}
  \state{Queensland}
  \country{Australia}
  \postcode{4000}
}
\author{Kien Nguyen*}
\email{k.nguyenthanh.edu.au}
\affiliation{%
  \institution{Queensland University of Technology}
  \streetaddress{2 George Street}
  \city{Brisbane}
  \state{Queensland}
  \country{Australia}
  \postcode{4000}
}
\author{Clinton Fookes}
\email{c.fookes@qut.edu.au}
\affiliation{%
  \institution{Queensland University of Technology}
  \streetaddress{2 George Street}
  \city{Brisbane}
  \state{Queensland}
  \country{Australia}
  \postcode{4000}
}
\author{George Karniadakis}
\email{george\_karniadakis@brown.edu}
\affiliation{%
  \institution{Brown University}
  \streetaddress{69 Brown St.}
  \city{Providence}
  \state{Rhode Island}
  \country{USA}
  \postcode{02912}
}

\thanks{*Corresponding author}

\renewcommand{\shortauthors}{Banerjee et al.}

\begin{abstract}
The incorporation of physical information in machine learning frameworks is opening and transforming many application domains. Here the learning process is augmented through the induction of fundamental knowledge and governing physical laws. In this work, we explore their utility for computer vision tasks in interpreting and understanding visual data. We present a systematic literature review of more than 250 papers on formulation and approaches to computer vision tasks guided by physical laws. We begin by decomposing the popular computer vision pipeline into a taxonomy of stages and investigate approaches to incorporate governing physical equations in each stage. Existing approaches in computer vision tasks are analyzed with regard to what governing physical processes are modeled and formulated, and how they are incorporated, i.e. modification of input data (observation bias), modification of network architectures (inductive bias), and modification of training losses (learning bias). The taxonomy offers a unified view of the application of the physics-informed capability, highlighting where physics-informed learning has been conducted and where the gaps and opportunities are. Finally, we highlight open problems and challenges to inform future research. While still in its early days, the study of physics-informed computer vision has the promise to develop better computer vision models that can improve physical plausibility, accuracy, data efficiency, and generalization in increasingly realistic applications. 
\end{abstract}


\maketitle

\section{Introduction}
Recent computer vision advancements have achieved exceptional performance in tasks like image classification, object detection, and human pose estimation \cite{DLsurvey}. Yet, these achievements often rely on complex, data-intensive models lacking robustness, interpretability, and alignment with physical laws and commonsense reasoning \cite{XAI,AdvAttacks}. Real-world phenomena, such as human motion and fluid dynamics, are governed by physical laws like Navier Stokes equations and constraints on anatomical movement \cite{zhang2021spatiotemporal,zhang2021three,chen2022physics,NERFVAE,gartner2022trajectory,gartner2022differentiable}, highlighting a gap in current approaches. Humans intuitively apply these physical principles for more effective interaction with the environment \cite{karniadakis2021physics,hao2022physics}. Physics-based methods, grounded in fundamental equations and domain knowledge, promise improved reliability and system safety by embodying the actual physical relationships at play \cite{balageas2010structural,hu2019case}, suggesting a need for a paradigm shift towards incorporating physical laws in computer vision.

Recent studies highlight the advantages of incorporating physics principles with machine learning, establishing a dominant paradigm in the field. Physics-informed machine learning (PIML), which integrates mathematical physics into machine learning models, enhances solution relevance and efficiency. This approach accelerates neural network training, improves model generalization with less data, and manages complex applications while ensuring solutions adhere to physical laws \cite{karniadakis2021physics,hao2022physics}.
Incorporating physical principles into machine learning, as seen in PI approaches, significantly boosts the robustness, accuracy, efficiency, and functionality of computer vision models \cite{hao2022physics,meng2022physics,karniadakis2021physics}. Visual data, including images, videos, and 3D point clouds, display complex characteristics that require domain-specific physics knowledge for effective processing, setting them apart from 1D signals. This distinction underlines the need for models specifically designed for computer vision tasks, leading to the exploration of the PICV field. The paper reviews state-of-the-art physics-informed strategies in computer vision, focusing on how physics knowledge is integrated into algorithms, the physical processes modeled as priors, and the specialized network architectures or augmentations employed to weave in physics insights.
\vspace{0.2cm}

\begin{figure}[t]
\centering
\def\arraystretch{1.6}%
\scalebox{0.8}{
\begin{tabular}{cc}
\subf{\includegraphics[width=0.47\columnwidth]{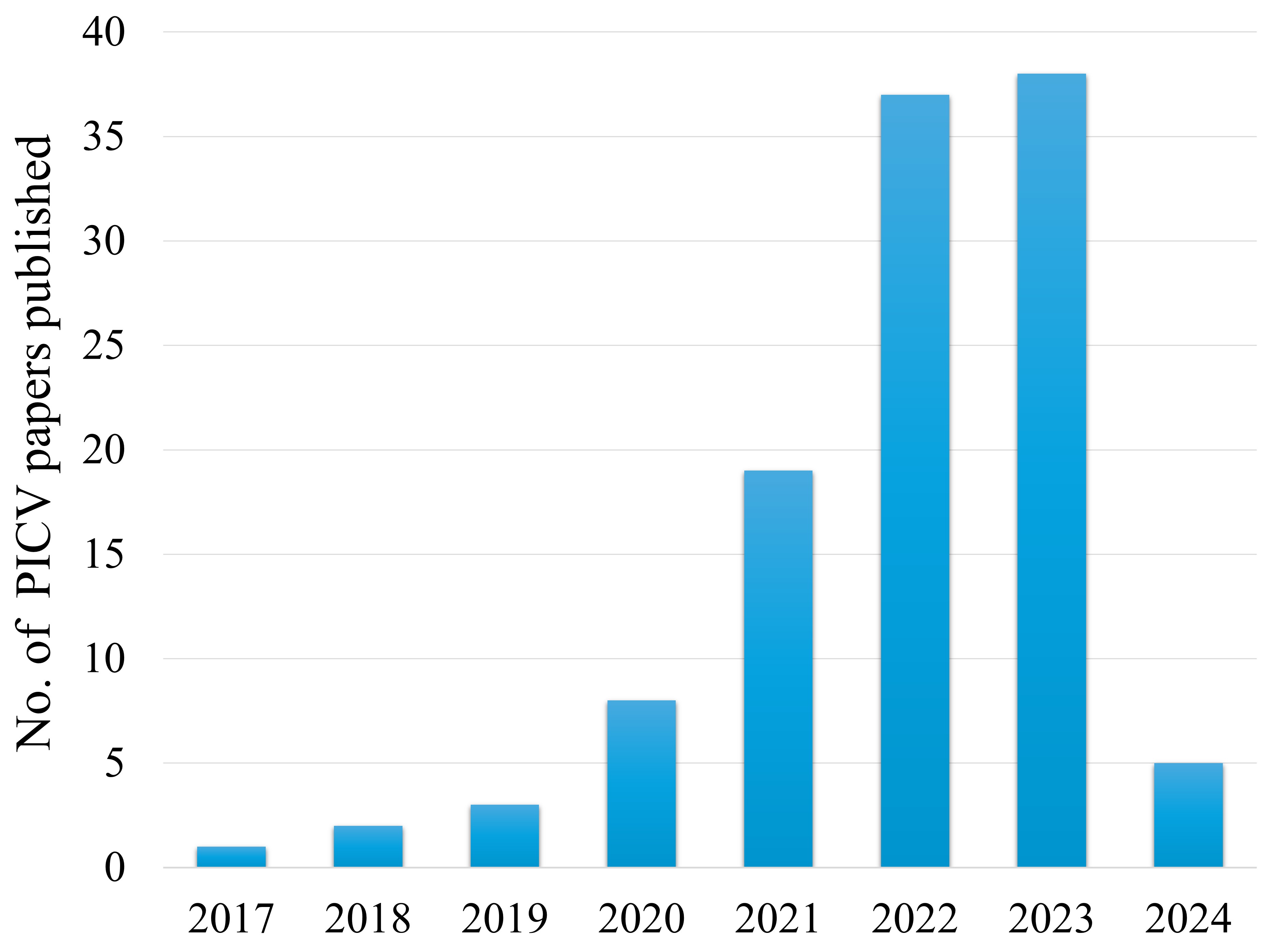}}
     {(a) PICV papers published over years}
&
\subf{\includegraphics[width=0.7\columnwidth]{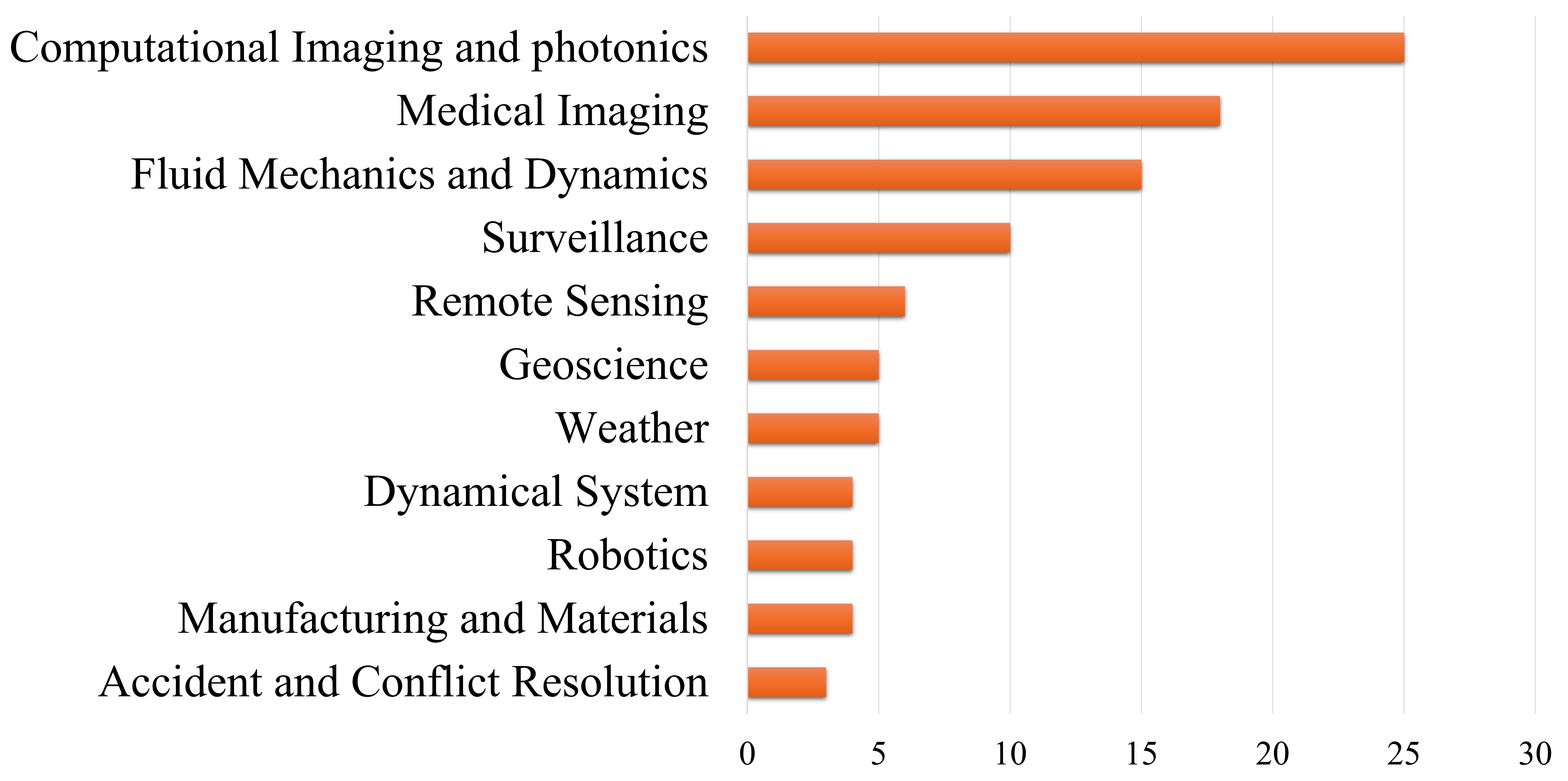}}
     {(b) Application domains of recent PICV papers}
\\
\end{tabular}}
\caption{ \textbf{(a)} Timeline of PICV papers published over the last eight years, where the histogram presents an exponentially increasing trend,  \textbf{(b)} Application domains of recent PICV papers. The most applied domain is computational imaging and photonics, closely followed by medical imaging.}
\label{fig:PICV timeline and application statistics}
\end{figure}

PICV is an increasing trend as illustrated in the increasing number of papers published in this area over the last 8 years, see Fig.~\ref{fig:PICV timeline and application statistics}a. The bar chart suggests that growing attention has been paid to this burgeoning field and we can expect many more to come.

Our contributions in this paper are summarized as follows:
\begin{itemize}
    \item We propose a unified taxonomy to investigate what physics knowledge/processes are modeled, how they are represented, and the strategies to incorporate them into computer vision models.
    \item We delve deep into a wide range of computer vision tasks, from imaging, super-resolution, generation, forecasting, and image reconstruction, to image classification, object detection, image segmentation, and human analysis. 
    \item In each task, we review in detail how physics information is integrated into specific computer vision algorithms for each task category, what physical processes have been modeled and incorporated, and what network architectures or network augmentations have been utilized to incorporate physics. We also analyze the context and datasets employed within these tasks.
    \item Based on the review of tasks, we summarize our perspectives on the challenges, open research questions, and directions for future research. 

 We discuss some open problems w.r.t. PICV, e.g., choosing the proper physics prior and developing a standard benchmarking platform. We also point out that tasks like human tracking, object detection, and video analysis have yet to leverage physics prior completely and thus have a vast space for research.  
\end{itemize}

\noindent
\textbf{Differences to other survey paper}:\newline
The field of physics-informed machine learning (PIML) is rapidly expanding, highlighted by surveys across various domains \cite{hao2022physics} including cyber-physical systems \cite{rai2020driven}, hydrology \cite{zhang2019recent}, fluid mechanics \cite{cai2022physics}, and climate modeling \cite{kashinath2021physics}. Specialized reviews have also focused on areas like medical imaging \cite{liu2021anatomy,wijesinghe2021emergent} and crowd analysis \cite{zhang2018physics}, which are pertinent to our broader computer vision scope. Our survey extends these efforts by offering a comprehensive view, identifying established areas, and underscoring emerging opportunities within physics-informed computer vision (PICV). {\color{black}Data for this review was systematically collected from major academic databases, including IEEE Xplore, ACM Digital Library, and others, emphasizing peer-reviewed journals and conference proceedings.}

The paper is structured as follows: Section~\ref{sec:PICV} introduces a taxonomy for integrating physics into computer vision models across various stages of the computer vision pipeline. Section~\ref{sec:PICVtasks} explores specific task groups within PICV, such as imaging, generation, super-resolution, and more.
Section~\ref{sec: quant_analysis} provides a quantitative analysis of the performance enhancements in CV tasks achieved through PI incorporation, and discusses key insights concerning its integration. Challenges and future research directions are discussed in Section~\ref{sec:openresearch}, with concluding remarks in Section~\ref{sec:conclusion}.


\section{Physics-informed Computer Vision: background, taxonomy, and examples}
\label{sec:PICV}
This section outlines a unified taxonomy of the integration of physics principles into computer vision models. Initially, we introduce the concept of PIML. Following this, we explore the application within computer vision, using a computer vision pipeline to illustrate the injection points and methods of incorporating physics into these models. Finally, we examine the practical uses of PICV models.

\subsection{Physics-informed Machine Learning (PIML)}\label{PIML_intro}

PIML aims to integrate mathematical physics models and observational data into the learning process to guide it towards physically consistent solutions in scenarios that are partially observed, uncertain, and high-dimensional \cite{kashinath2021physics,hao2022physics,cuomo2022scientific}. Including physics information, which represents the fundamental principles of the process being modeled, enhances ML models by providing significant advantages \cite{kashinath2021physics,meng2022physics}.

\begin{enumerate}
    \item Makes the ML model both physically and scientifically consistent.
    \item Model training becomes highly data-efficient, i.e. trainable with fewer data.
    \item Accelerates the model training process, such that the models converge faster to an optimal solution.
    \item Makes the trained models highly generalizable, such that models can make better predictions for scenarios unseen during the training phase.
    \item Improves transparency and interpretability of models thus making them explainable and more trustworthy.
\end{enumerate}




\noindent Conventional literature has shown three strategies to incorporate physics knowledge/priors into machine learning models: observational bias, learning bias, and inductive bias. 


\textbf{Observational bias:} It utilizes multi-modal data, which is expected to reflect the underlying physical principles which dictate their generation \cite{lu2021learning,kashefi2021point,li2020fourier,yang2019conditional}. The underlying deep neural network (DNN) is exposed directly to the training/ observed data and the DNN is expected to capture the underlying physical process via training. The training data seen by the DNN can come from direct observations, simulation/ physical equation-generated data, maps, and extracted physics data induction.

\textbf{Learning bias:} enforces prior knowledge/ physics information through soft penalty constraints. Approaches in this category augment loss functions with additional terms that are based on the physics of the underlying process, e.g. momentum, conservation of mass, etc. For example, physics-informed neural networks (PINN) integrate the information from both the measurements and partial differential equations (PDEs) by embedding the PDEs into the loss function of a neural network using automatic differentiation \cite{karniadakis2021physics}. 
Some prominent examples of soft penalty-based approaches include statistically constrained GAN \cite{wu2020enforcing}, PI auto-encoders \cite{erichson2019physics}, and encoding invariances by soft constraints in the loss function InvNet \cite{shah2019encoding}. 

\textbf{Inductive biases:} prior knowledge can be incorporated through custom neural network induced 'hard' constraints. 
For example, Hamiltonian NN \cite{greydanus2019hamiltonian} encodes better inductive biases to NNs, draws inspiration from Hamiltonian mechanics, and trains models such that they respect exact conservation laws. Cranmer et al. introduced Lagrangian Neural Networks (LNNs) \cite{cranmer2020lagrangian}, which can parameterize arbitrary Lagrangians using neural networks and unlike most HNNs, LNNs can work where canonical momenta are unknown or difficult to compute. 
{\color{black}Besides tensor basis networks (TNNs) \cite{ling2016reynolds}  incorporate tensor algebra into their operations and structure, allowing them to exploit the high-dimensional structure of tensor data more effectively than traditional neural networks.}
\cite{meng2022learning} uses a Bayesian framework where functional priors are learned using a PI-GAN from data and physics. Followed by using the Hamiltonian Monte Carlo (HMC) method to estimate the posterior PI-GAN's latent space. It also uses special DeepONets \cite{lu2021learning} networks in PDE agnostic physical problems.

\subsection{Physics-Informed Computer Vision (PICV)}
{\color{black}

\subsubsection{Physics incorporation in general ML and CV:} 
Incorporating physics knowledge in CV tasks offers unique benefits that differ from those of general ML tasks. 
General ML encompasses a wide range of applications, including natural language processing and predictive analytics, while the use of physics knowledge is typically limited to tasks that are based on physical phenomena, such as predicting weather patterns or simulating physical systems.
Many ML tasks rely heavily on data-driven methods to learn patterns from datasets. However, the use of physics knowledge can improve model development and interpretability by incorporating underlying physical principles.
Feature engineering is critical for model performance in ML tasks. PI aids in feature selection and creation by incorporating known causal relationships, particularly for time-series and spatial data predictions.

In CV, tasks often require the ability to understand and analyze spatial and temporal dynamics in visual data. Physics can be explicitly used to model and predict physical behaviors and interactions within images and videos, such as object motion, light propagation, and material properties, for instance in image reconstruction.
Physics-based constraints and regularization techniques, such as geometric constraints and conservation laws, are often used in CV to guide the learning process. This ensures that the solutions are physically compliant, which is particularly useful in cases where visual data alone is ambiguous or insufficient.
CV tasks can greatly benefit from the integration of multiple data modalities to achieve a comprehensive understanding of the scenes. In this process, physics plays a crucial role in aligning and fusing data from different sensors.
}

\begin{figure}[h]
  \centering
  \includegraphics[width=0.6\columnwidth]{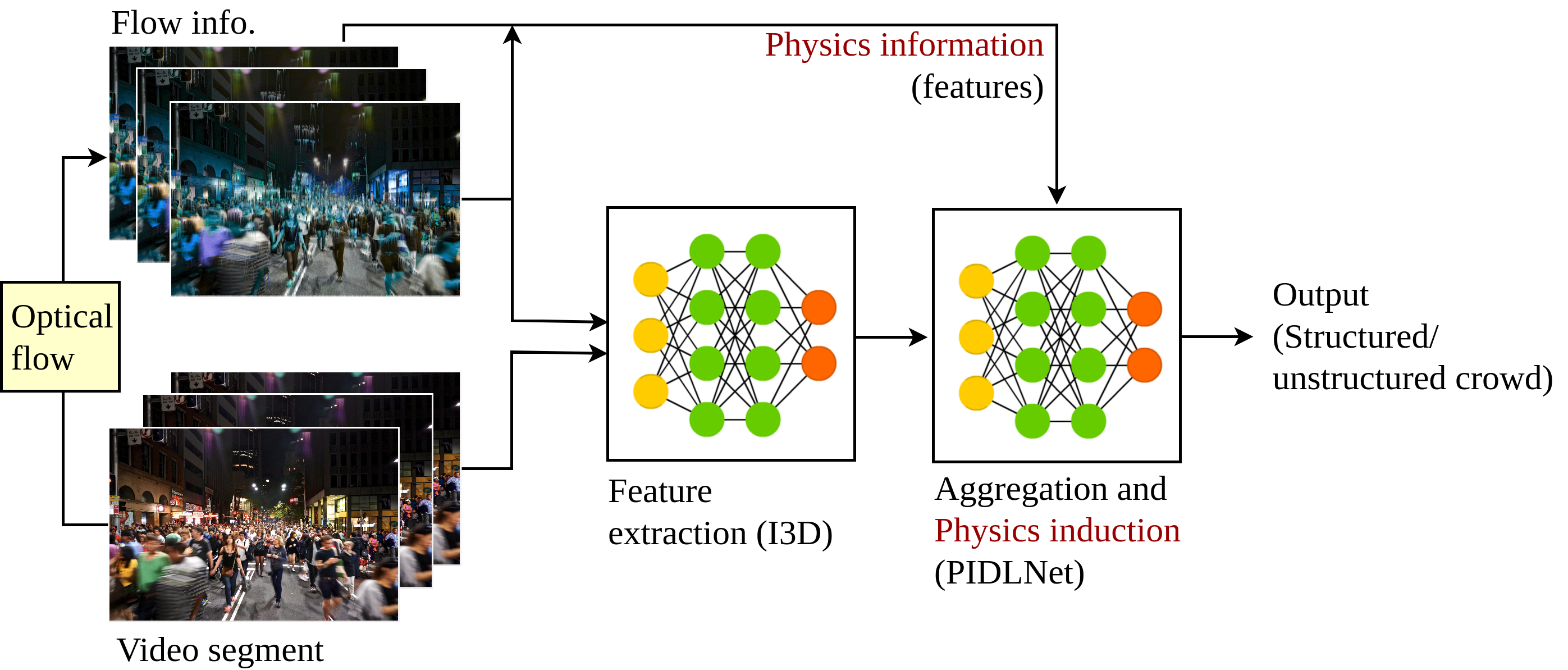}
  \caption{A simplified illustrative example of physics incorporation in a computer vision task, adapted from \cite{behera2021pidlnet}. Physics information, in the form of flow data, is extracted from video sequences and incorporated into an aggregating network (PIDLNet).}
  \Description{}
  \label{fig:intuit_phy_example}
\end{figure}

\subsubsection{ Intuitive introduction to physics priors in CV:}   
Several intuitive physical rules/ constraints have been efficiently leveraged in CV tasks. For example, in the task of human analysis, works use prior knowledge about the biological structure of the human body (e.g., arms, head, and legs are connected to the torso)\cite{isogawa2020optical} and anatomical body joint limits \cite{gartner2022differentiable}. This physics incorporation ensures compliance of the solutions to the physical plausibility of human structure and motion. Other constraints may include contacts \cite{livne2018walking}, temporal consistency, and collision. On similar lines a number of works especially in human analysis have substantially used human dynamics models or physics simulators to generate pose references for tasks like motion estimation/ generation \cite{yuan2022physdiff,zhang2022pimnet}, motion capture \cite{huang2022neural} and 3D pose estimation \cite{yuan2021simpoe}. In other words where physical variables form part of the overall loss function, domain knowledge-based intuition is of special significance. E.g. in \cite{li2022using}, authors introduce an additional physics-based constraint in the loss function, based on the intuition that along with the traditional MSE term, the objective should also include the difference of the volume of liquid phase between the input and the output, in this super-resolution task concerned with fluid flow.
In \cite{behera2021pidlnet} the authors introduce a framework that is trained on both conventional and two physics-based features: order and entropy, for the characterization of crowd movement as structured and unstructured. Drawing intuition from physics, a low entropy and unity order can be attributed to ordered crowd movement. While high entropy and order parameter values signify random pedestrian movement and that movement is highly curved, respectively.
These parameters are obtained from the motion flows extracted from the crowd videos, and later coupled with the aggregated output, see Fig~\ref{fig:intuit_phy_example}.

\begin{figure}
\centering
\def\arraystretch{2}%
\scalebox{0.9}{
\begin{tabular}{ll}
\subf{\includegraphics[width=0.7\columnwidth]{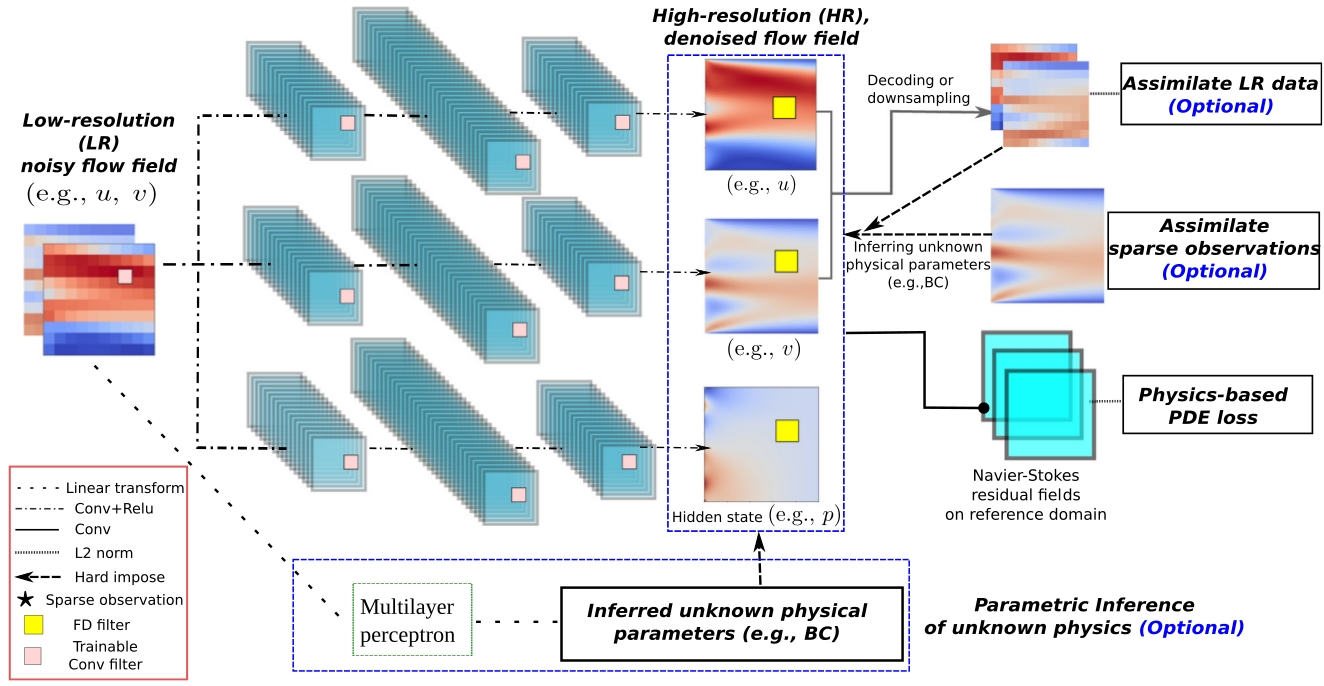}}
     {(a) PDE as physics prior \cite{gao2021super} }
     \label{fig:example_diff_eqn}
&
\subf{\includegraphics[width=0.35\columnwidth]{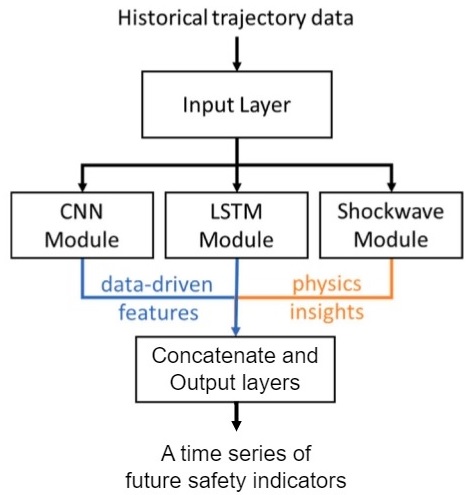}}
     {(b) Historical data as physics-prior \cite{yao2023physics}}
     \label{fig:exampple_Hist_data}
\\
\subf{\includegraphics[width=0.65\columnwidth]{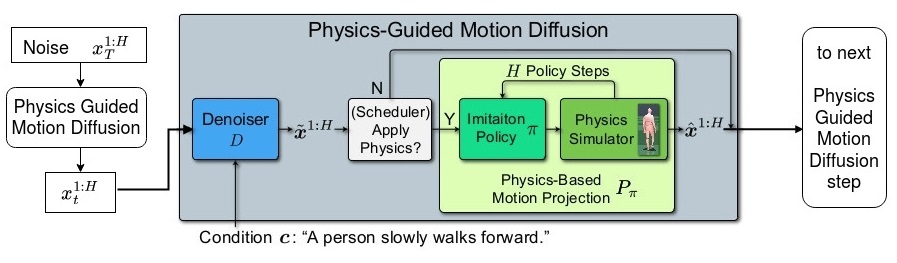}}
     {(c) Physics model as physics-prior \cite{yuan2022physdiff}}
     \label{fig:example_sim}
&
\subf{\includegraphics[width=0.35\columnwidth]{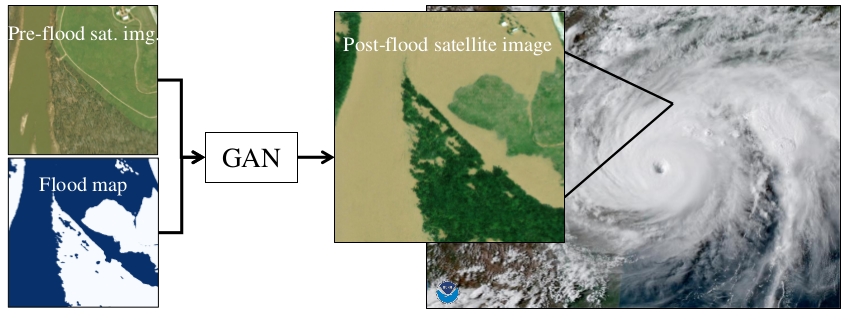}}
     {(d) Visual data as physics prior\cite{lutjens2020physics}}
     \label{fig:example_map}
\\
\subf{\includegraphics[width=0.65\columnwidth]{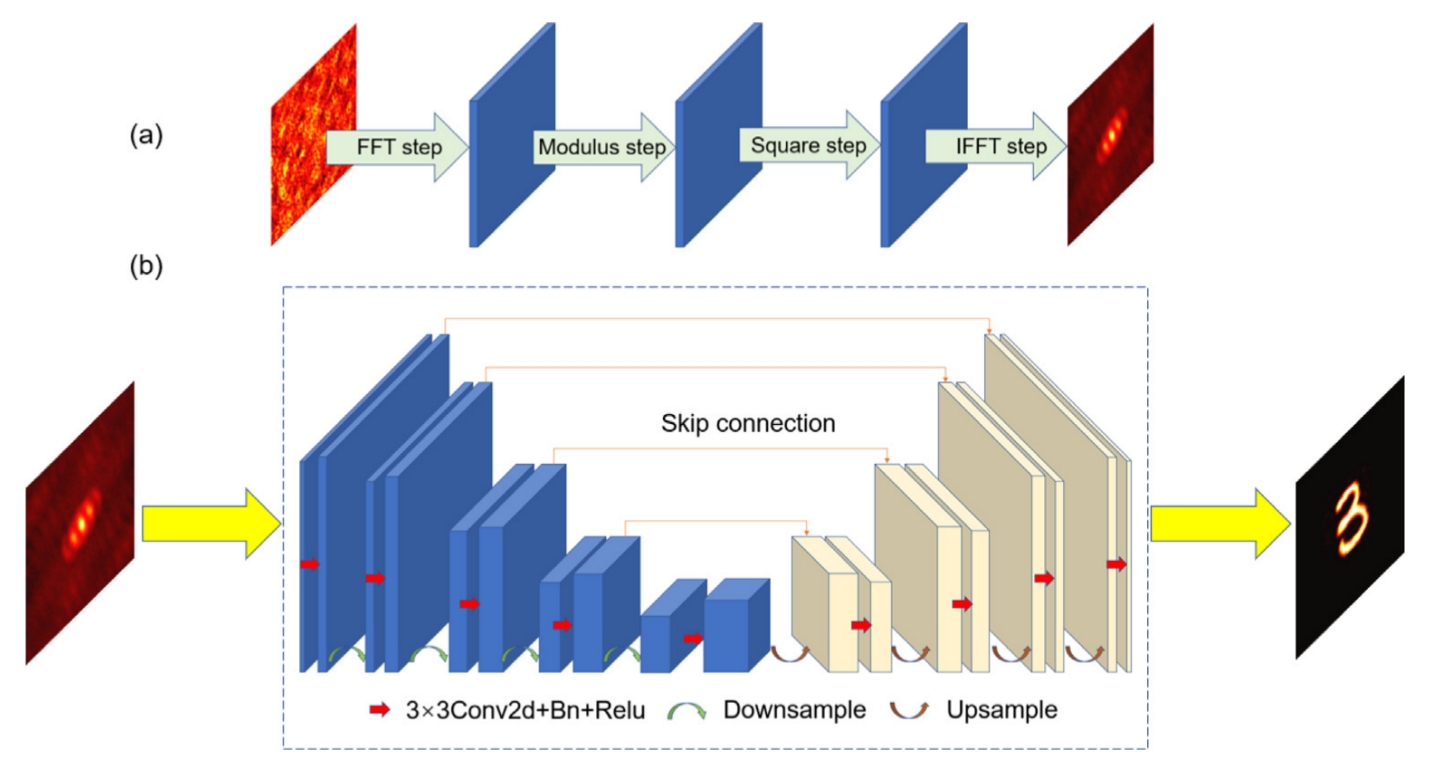}}
     {(e) Statistical property as physics-prior\cite{guo2023dynamic} }
     \label{fig:SR_example}
&
\subf{\includegraphics[width=0.37\columnwidth]{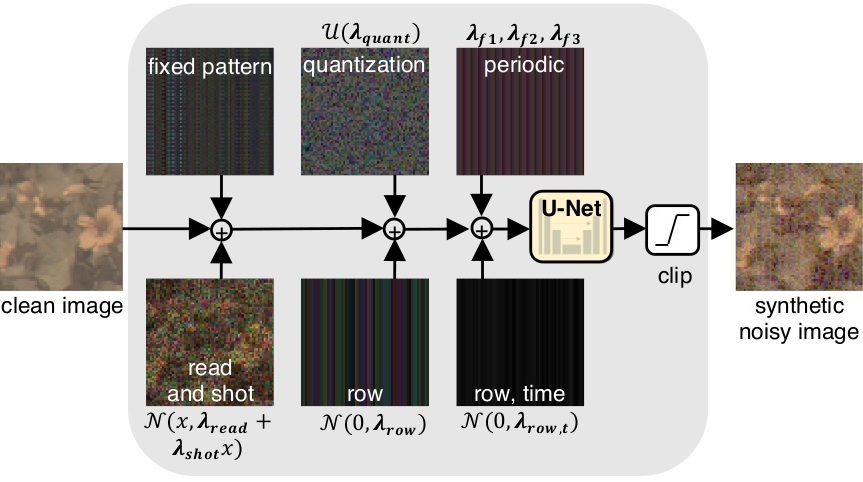}}
     {(f) Physical variable as physics-prior\cite{monakhova2022dancing} }
     \label{fig:example_comp}
\\
\end{tabular}}

\caption{ Different physics prior examples. For Governing eqns. and constraints type priors \textbf{(a)} PDE as physics prior \cite{gao2021super}; here a PDE loss is used to complement traditional network training and \textbf{(c)} Physics model as physics prior, \cite{yuan2022physdiff}; here a physics simulator is used for motion projection for generating physically-plausible human motions, \textbf{(b)} Physics via historical data  \cite{yao2023physics}; here historical trajectory data is used by deep network to derive physics insights and data-driven features,  \textbf{(d)} Physics information as visual representation \cite{lutjens2020physics}; here a GAN pipeline ingests flood maps as physics prior along-with pre-flood satellite images generating photorealistic post-flood images, \textbf{(e)} Physics information as statistical property \cite{guo2023dynamic}; here using speckle redundancy, the speckles from different configurations are described by different sub-regions of speckles from a single configuration. Such pre-processed speckle pattern is fed to NN post-processing module for object reconstruction, \textbf{(f)} Physics information as physical variable \cite{monakhova2022dancing}; here a generative noise model (UNet) is based on physical noise parameters, where these parameters are based on prior knowledge of random variable distributions which can approximately model these noise types.}

\label{fig:Physics prior types}
\end{figure}

\begin{figure}[h]
\centering
\def\arraystretch{2}%
\scalebox{0.7}{
\begin{tabular}{ll}
\subf{\includegraphics[width=0.85\columnwidth]{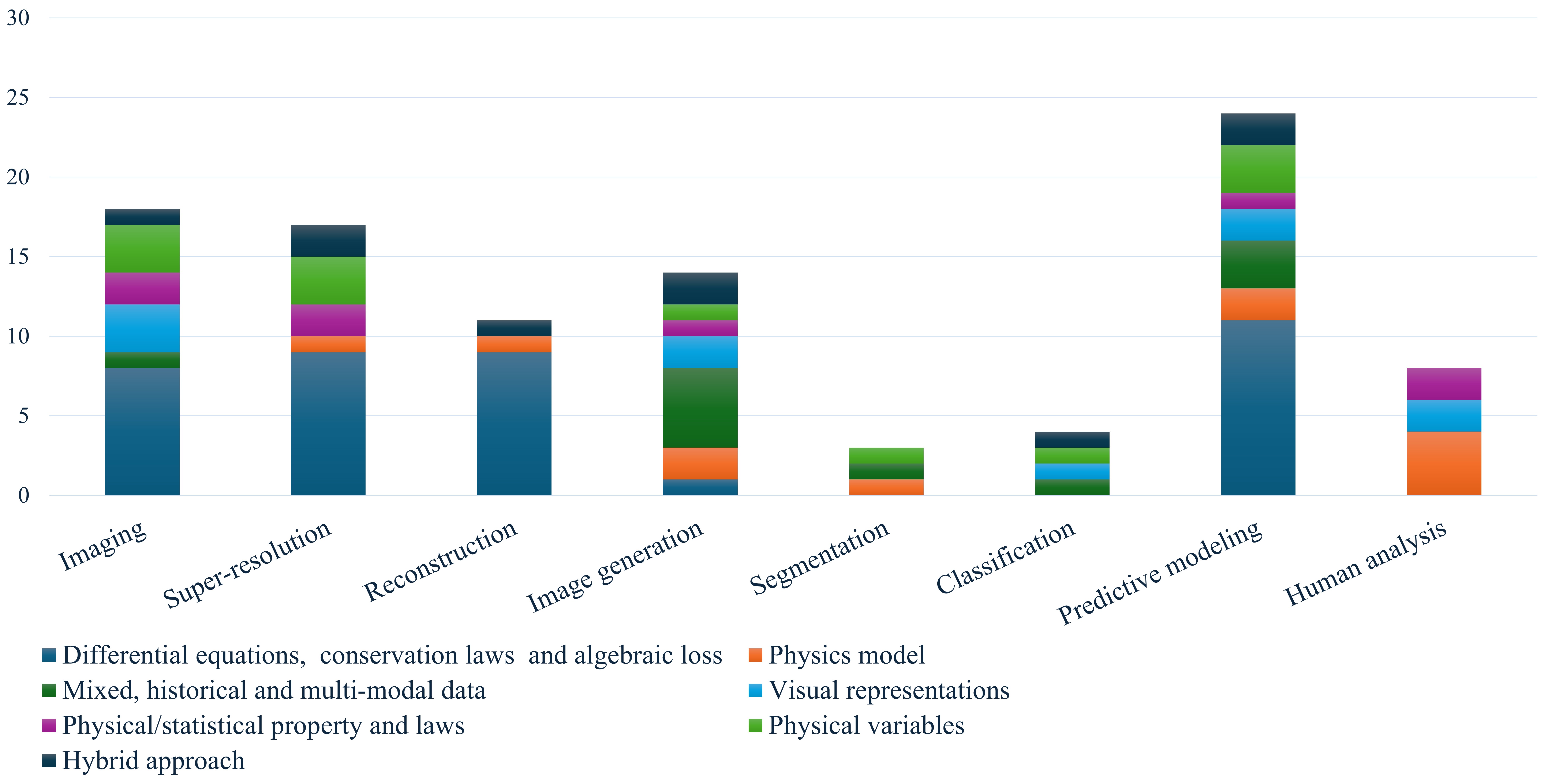}}
     {(a) Types of physics priors used in each CV task.}
&
\subf{\includegraphics[width=0.55\columnwidth]{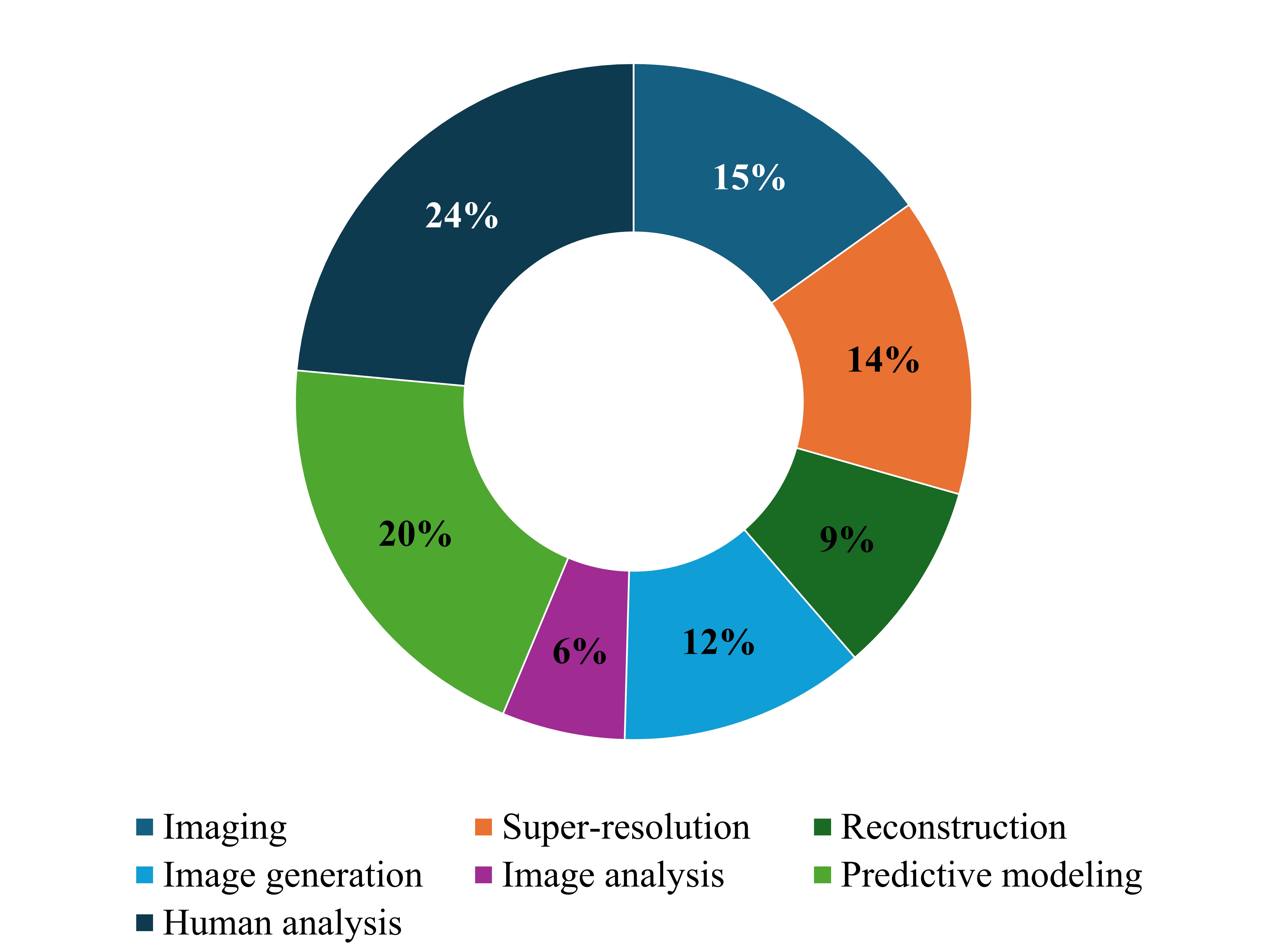}}
     {(b) Research share of each CV task.}
\\
\end{tabular}}
\caption{ \textbf{(a)} The stacked histogram presents the statistics of a certain type of physics prior in a specific CV task, \textbf{(b)} The pie chart presents the research share of PI approaches in different CV tasks. Task ``Image analysis'' constitutes classification and segmentation tasks.}
\Description{}
\label{fig:PICV_prior_types_and_research_share}
\end{figure}

\begin{table}[h!]
\caption{Categorization of latest PICV papers with regards to type of physics priors}
\centering
\scalebox{0.6}{
\begin{tabular}{|l|l|l|l|l|l|l|l|}
\hline 
\multicolumn{6}{|c}{\textbf{Physics information types }} & & \textbf{Computer vision task}  \\
\hline
Differential equations, & Physics model & Mixed, historical & Visual & Physical/ statistical  & Physical  & Hybrid & - \\
conservation laws and algebraic loss & 
& 
and multi-modal data &
representations &
property and laws &
variables & 
approach &\\
\hline
\cite{chen2022physics,zhang2020physics,saba2022physics,eichhorn2024physics,zhu2023physics,yang2023fwigan,kamali2023elasticity,halder2023mri}& 
& 
\cite{deng2020interplay} & 
\cite{monakhova2019learned,yanny2022deep,poirot2019physics}& 
\cite{guo2023dynamic,zhu2021imaging} & 
\cite{monakhova2022dancing,weiss2019pilot,xypakis2022physics} & 
\cite{bian2023high} & 
\textbf{Imaging} \\

\hline

\cite{kelshaw2022physics,arora2022spatio,eivazi2022physics,ren2022physics,fathi2020super,shu2023physics,shone2023deep,trinh20243d,fok2023deep} & 
& 
\cite{yasuda2022super}& 
& 
\cite{subramaniam2020turbulence,li2023learning} & 
\cite{bode2021using,chen2023physics,pradhan2023variational} & 
\cite{zayats2022super,haghighi2023accelerating} & 
\textbf{Super-resolution } \\

\cite{li2022using,wang2020physics,arora2022physrnet} & & & & & & &\\
\hline

\cite{shu2023physics,zhang2021three,wang2022dense,molnar2023estimating,chu2022physics,ning2023image,delcey2023physics,mondal2024physics,burns2023untrained}& 
\cite{yin2024physics}& 
& 
& 
& 
& 
\cite{chen2022aug} & 

\textbf{Reconstruction}   \\

\hline

\cite{zheng2020physics}& 
\cite{Pizzati2023PIGAN,yang2023physics} & 
\cite{siddani2021machine,qian2022physics,manyar2023physics,kawahara2023mri,pan20232d} & 
\cite{lutjens2020physics,liu2022physics}& 
\cite{oviedo2019fast} & 
\cite{thoreau2022p}& 
\cite{chen2021physics,borges2024acquisition} & 
\textbf{Image generation}   \\

\hline

& 
\cite{jenkins2020physics} & 
\cite{borges2019physics} & 
& 
& 
\cite{cciccek20163d} & 
& 
\textbf{Segmentation}   \\
\hline

& 
& 
\cite{altaheri2022physics} & 
\cite{guc2021fault}& 
& 
\cite{du2021physics} & 
\cite{lai2020full} & 
\textbf{Classification}   \\
\hline

\hline
\cite{kissas2020machine,lipac,ni2022ntfields,zapf2022investigating,herrero2022ep,van2022physics,zhao2023physics,lopez2023warppinn}
& 
\cite{muller2022deep,ma2022risp} & 
\cite{yao2023physics,chen2022deepurbandownscale}& 
\cite{zhou2021harnessing,zhao2021physics} & 
\cite{sarabian2022physics} & 
\cite{arun2023physics,garttner2021estimating,wu2018seeing} & 
\cite{zhang2021spatiotemporal,mehta2020physics} 
& 
\textbf{Predictive modeling} \\

\cite{sahli2020physics,oikonomou2022physics,cai2021flow} & 
& 
\cite{zantedeschi2020towards,fablet2021end,yang2021revisit}& 
& 
& 
& 
\cite{wei2022fracture,buoso2021personalising}& 
\\

\hline

& 
\cite{gartner2022differentiable,gartner2022trajectory,isogawa2020optical,xu2023interdiff}& 
& 
& 
\cite{livne2018walking,behera2021pidlnet} & 
\cite{murray2017bio,xie2021physics}  & 
& 
\textbf{Human analysis}   \\

& 
\cite{yuan2021simpoe,zhang2022pimnet,huang2022neural,yuan2022physdiff,yi2022physical}& 
& 
& 
& 
& 
& 
\\
\hline

\end{tabular}}
\label{table:PIML_CV_categorisation}
\end{table}

\subsubsection{Physics prior categories with examples:} Based on the source of the physics information they can be categorized in the following typical categories,  as presented in Fig~\ref{fig:Physics prior types}.
{\color{black} \textit{Governing equations and constraints} category, leverage differential equations and physical laws, ensuring predictions adhere to foundational principles, crucial for tasks requiring physical accuracy. Differential equations, conservation laws, and algebraic loss specifically embed these laws into models, like PINNs, for tasks like super-resolution, where physical realism is paramount. Physics models use complete simulations for tasks demanding dynamic understanding, such as human motion analysis. 
\textit{Mixed, historical, and multi-modal data } approaches combine visual and empirical data with theoretical insights, enriching models with a comprehensive physical perspective. \textit{Visual representations} methods incorporate physics through varied visual data forms, aiding tasks with spatial and temporal physical dynamics. The \textit{physical/statistical property} category taps into fundamental system behaviors, offering a nuanced integration of core principles. \textit{Physical variables } include relevant parameters directly in models, enhancing predictions' physical consistency. Lastly, \textit{hybrid approaches} amalgamate these strategies, optimizing model performance by balancing empirical accuracy and physical law adherence, demonstrating a comprehensive, multifaceted approach to embedding physics in CV. }

A statistic on the different categories of physics priors used is provided in Fig.~\ref{fig:PICV_prior_types_and_research_share}a and Table~\ref{table:PIML_CV_categorisation}.

\begin{enumerate} 
    \item \textit{Governing equations and constraints: }
    {\color{black} We categorize approaches integrating governing equations and physical laws into the computer vision pipeline. Type A employs direct application of differential equations (DE, PDE) and physical constraints (conservation laws, symmetries), especially through PINN-based methods. Type B uses these mathematical descriptions to simulate physical phenomena, aiding in synthetic data generation and solving inverse problems, enhancing accuracy across tasks.
    
    \textit{A. Differential equations, conservation laws, and algebraic loss:} 
    A large number of works, leverage system dynamics representations in the form of partial/ordinary differential equations, as physics priors \cite{kelshaw2022physics,arora2022spatio,eivazi2022physics,ren2022physics}, especially through the use of PINN \cite{raissi2017physics} and suchlike special networks. PINNs assimilate information from measurement/ data as well as PDEs by incorporating the PDEs in the loss function of the neural network using automatic differentiation \cite{karniadakis2021physics}. Besides it is a common practice to constrain/ regularize the loss function using conservation laws of mass and/or momentum \cite{kissas2020machine,sarabian2022physics}. In certain papers e.g. \cite{li2022using}, an algebraic loss is also used.
    For example, \cite{gao2021super} in super-resolution CV task, produces high-resolution (HR) flow fields from low-resolution (LR) inputs in high-dimensional parameter space. The involved CNN-SR network is trained purely based on physical laws with strictly imposed boundary conditions and does not need HR data. See Fig.~\ref{fig:Physics prior types}a, which shows the inclusion of the PDE loss as part of the training paradigm.
    
    \textit{B. Physics model:} In a number of works a complete physics model has been used as a source of physics-based guidance for performing the CV task. Physics dynamics model \cite{zhang2022pimnet} and physics simulators \cite{isogawa2020optical,yuan2021simpoe,yuan2022physdiff} have been extensively used especially in human analysis tasks. For example, \cite{yuan2022physdiff} proposed a diffusion model that generates physically plausible human motions using a PI-motion projection module in the diffusion process. The said module uses motion imitation in a physics simulator for projecting the denoised motion of a diffusion step to a physically plausible motion, see Fig.~\ref{fig:Physics prior types}c. }
    \item {\color{black} \textit{Mixed, historical and multi-modal data: }
    In the mixed data-based approach \cite{manyar2023physics,siddani2021machine,isogawa2020optical}, the DNN is trained using both measurement data and data generated from physics-based models/ simulators. The goal here is to obtain a model which incorporates qualities from both the model and measurement data. In certain cases, data from past iterations or historical data have also been used as the source of physical information \cite{yao2023physics}, from which a physical concept is later learned by the networks. Multimodal data e.g. multi-spectral images do also serve as a source of physics information, e.g. in \cite{chen2022deepurbandownscale}, see Fig.~\ref{fig:Physics prior types}b.}

    \item \textit{Visual representations:} Physics information is also incorporated through different types of visual data, that by nature or through some processing on raw data contains physics information e.g. time-frequency signals \cite{guc2021fault}, maps \cite{lutjens2020physics} and hyper-spectral images \cite{wu2022intracity}. For example in \cite{lutjens2020physics}, a deep learning pipeline generates satellite images of current and future coastal flooding. A generative vision model learns physically-conditioned image-to-image transformation from pre-flood image to post-flood image, by leveraging physics information from flood extent map (mask) as input, see Fig.~\ref{fig:Physics prior types}d.
    %
    \item {\color{black} \textit{Physical/ statistical property: }} Physics information can also take the form of some physical or statistical property. For example, Shannon entropy is considered as physics information in \cite{deng2020interplay} and speckle correlation serves as physics information in \cite{guo2023dynamic}. In other cases \cite{oviedo2019fast}, physical property based on domain knowledge of the system has been leveraged, see Fig.~\ref{fig:Physics prior types}e. 
    \item \textit{Physical variables:} In this category physics information can come in the form of physically relevant variables which are either incorporated as additional data input to the CV model \cite{yasuda2022super,wu2018seeing,du2021physics} or as additional component(s) in the loss function used to train the CV model/ relevant network\cite{bode2021using,li2022using,monakhova2022dancing}. For example, in \cite{monakhova2022dancing} a generative noise model is designed to train a low light video denoiser, with PI statistical noise parameters, which are optimized during training to produce a synthetic noisy image that is indistinguishable from a real noisy image.  These noises are based on prior knowledge of random variable distributions which can approximately model these noise types, see Fig.~\ref{fig:Physics prior types}f.

    \item \textit{Hybrid approach: }
    In hybrid approaches, we include those works that have utilized combinations of any of the above categories. However in most cases \cite{wang2020physics,chen2021physics} the hybrid approach pairs simulated data with PI loss function, for better performance at CV tasks. 

\end{enumerate}

\subsubsection{Approaches to incorporate physics priors into computer vision models}

\begin{figure}
\centering
\def\arraystretch{1.8}%
\scalebox{1}{
\begin{tabular}{ll}
\subf{\includegraphics[width=0.48\columnwidth]{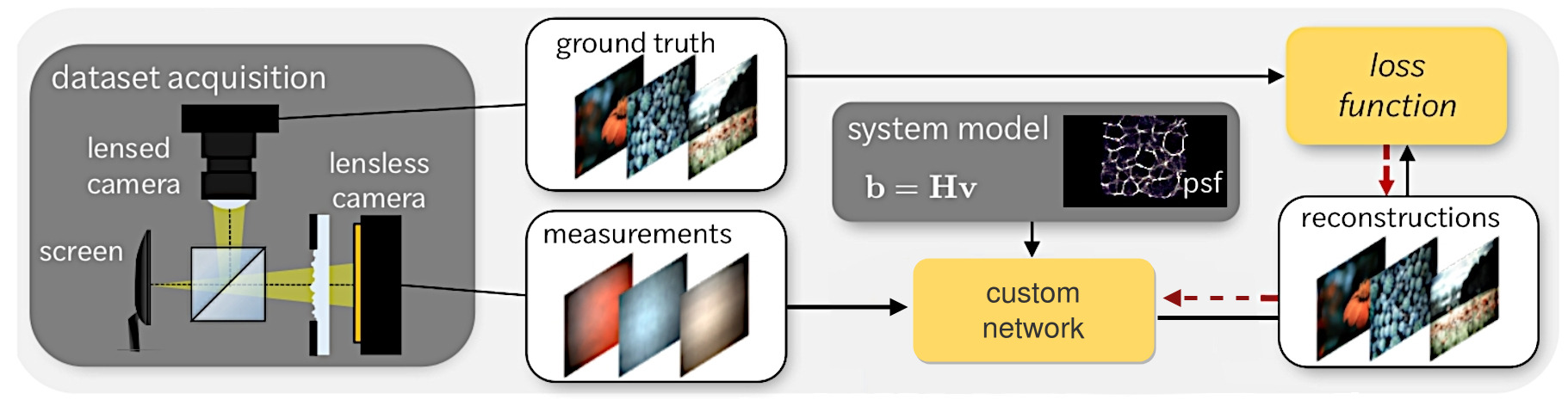}}
     {(a) Data acquisition stage}
     \label{fig:example_obs_bias}
&
\subf{\includegraphics[width=0.48\columnwidth]{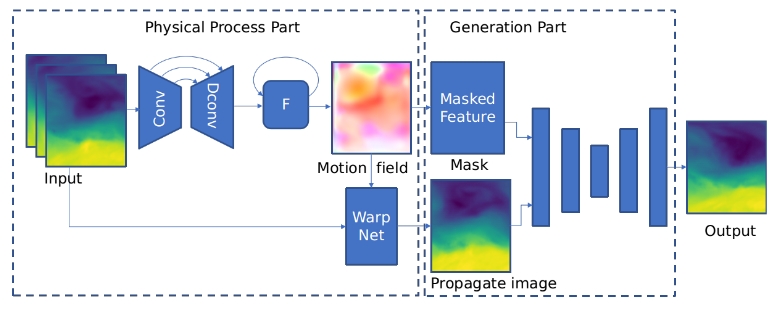}}
     {(b) Pre-processing stage }
     \label{fig:example_lrn_bias}
\\
\subf{\includegraphics[width=0.48\columnwidth]{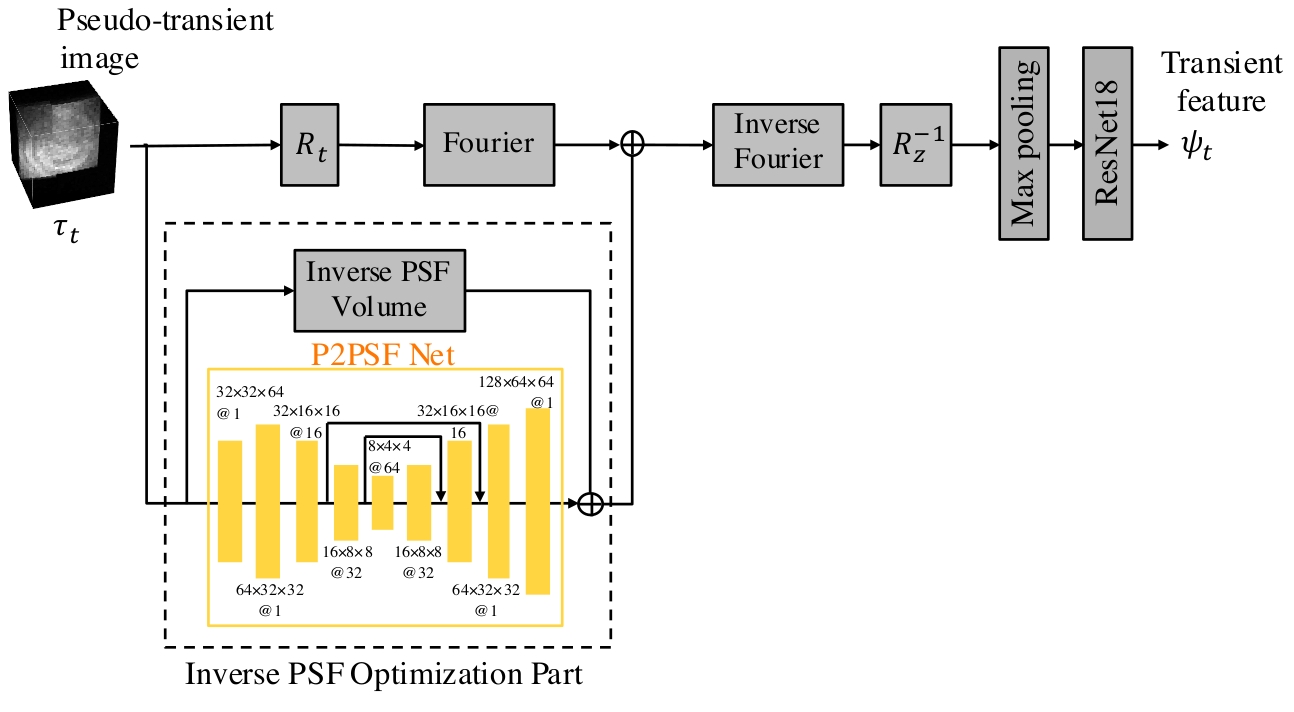}}
     {(c) Model design stage (feature extraction)}
     \label{fig:example_indc_bias}
&
\subf{\includegraphics[width=0.48\columnwidth]{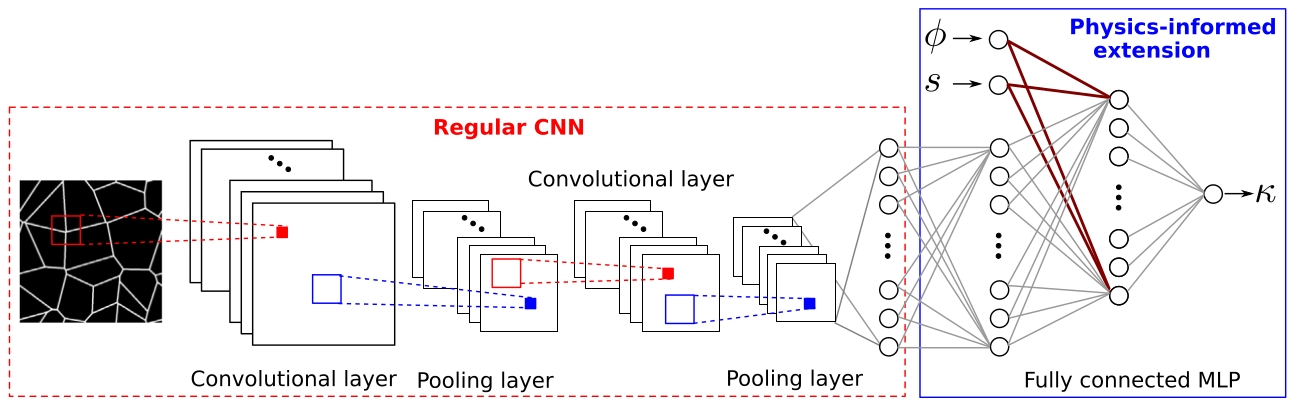}}
     {(d) Model design stage (architecture customization)}
     \label{fig:example_lrn_bias2}
\\ 
\subf{\includegraphics[width=0.48\columnwidth]{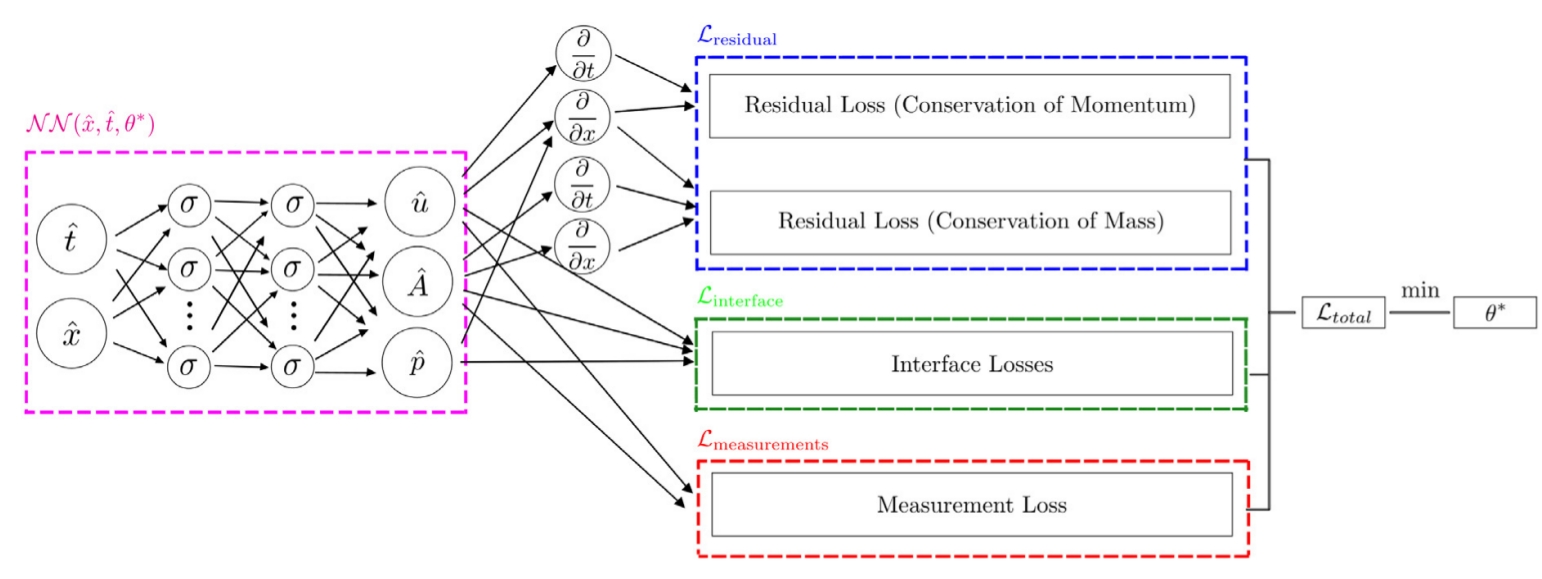}}
     {(e) Model training stage }
     \label{fig:example_lrn_bias_Pinn}
&
\subf{\includegraphics[width=0.48\columnwidth]{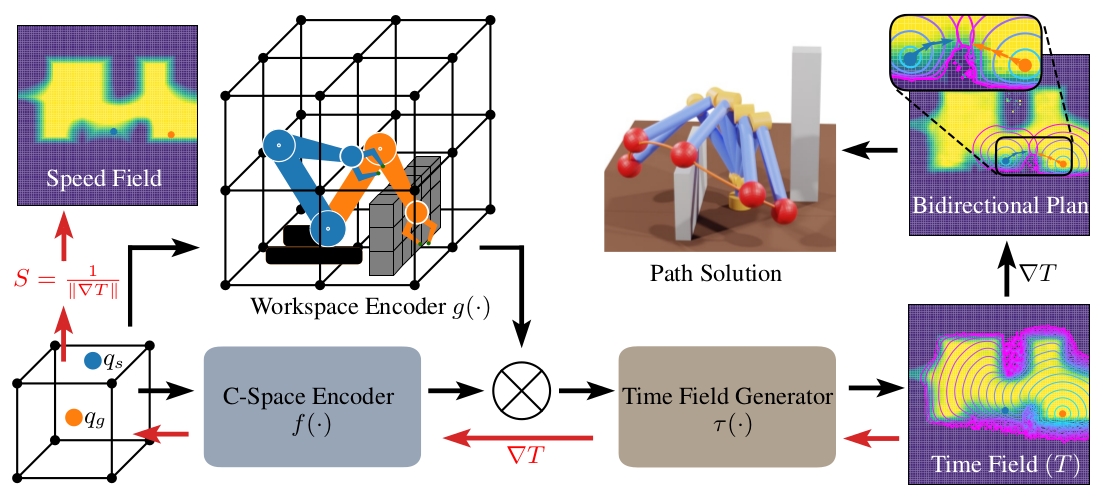}}
     {(f) Inference stage }
     \label{fig:example_inference}
\\
\end{tabular}}
\caption{Examples of physics incorporation with regard to the CV pipeline \textbf{(a)} Physics incorporation after data acquisition  \cite{monakhova2019learned}; in this imaging task the physics prior in the form of a physics system model is introduced to the custom NN after data acquisition, \textbf{(b)} Physics incorporation during image pre-processing \cite{chen2021physics}; in this temperature field generation task, the physical process module directly generates a motion field from input images and function (F) learns dynamic characteristics of the motion field, \textbf{(c)} Physics incorporation at model design (feature extraction) stage \cite{isogawa2020optical}; in this human analysis task, custom network (P2PSF net) is designed to extract transient feature from images, to model physically-consistent 3D human pose, \textbf{(d)} Physics incorporation at model design (architecture selection/ customization) stage \cite{wu2018seeing}, here in the PI extension of a regular CNN network, physical parameters are included during training for faster permeability prediction, \textbf{(e)} Physics incorporation at model training stage \cite{kissas2020machine}, in this prediction task \textbf{(f)} Shows end-to-end pipeline of a robot motion planning, which is also a CV prediction task, with the inference or end product being the path solution. The approach uses a physics-driven objective function and reflects it through the architecture to parameterize the PDE (Eikonal equation) and generate
time fields for different scenarios.} 
\label{fig:CVpipe_phy_incorp_example}
\end{figure}
\begin{figure}[t]
  \centering
  \includegraphics[width=0.7\linewidth] {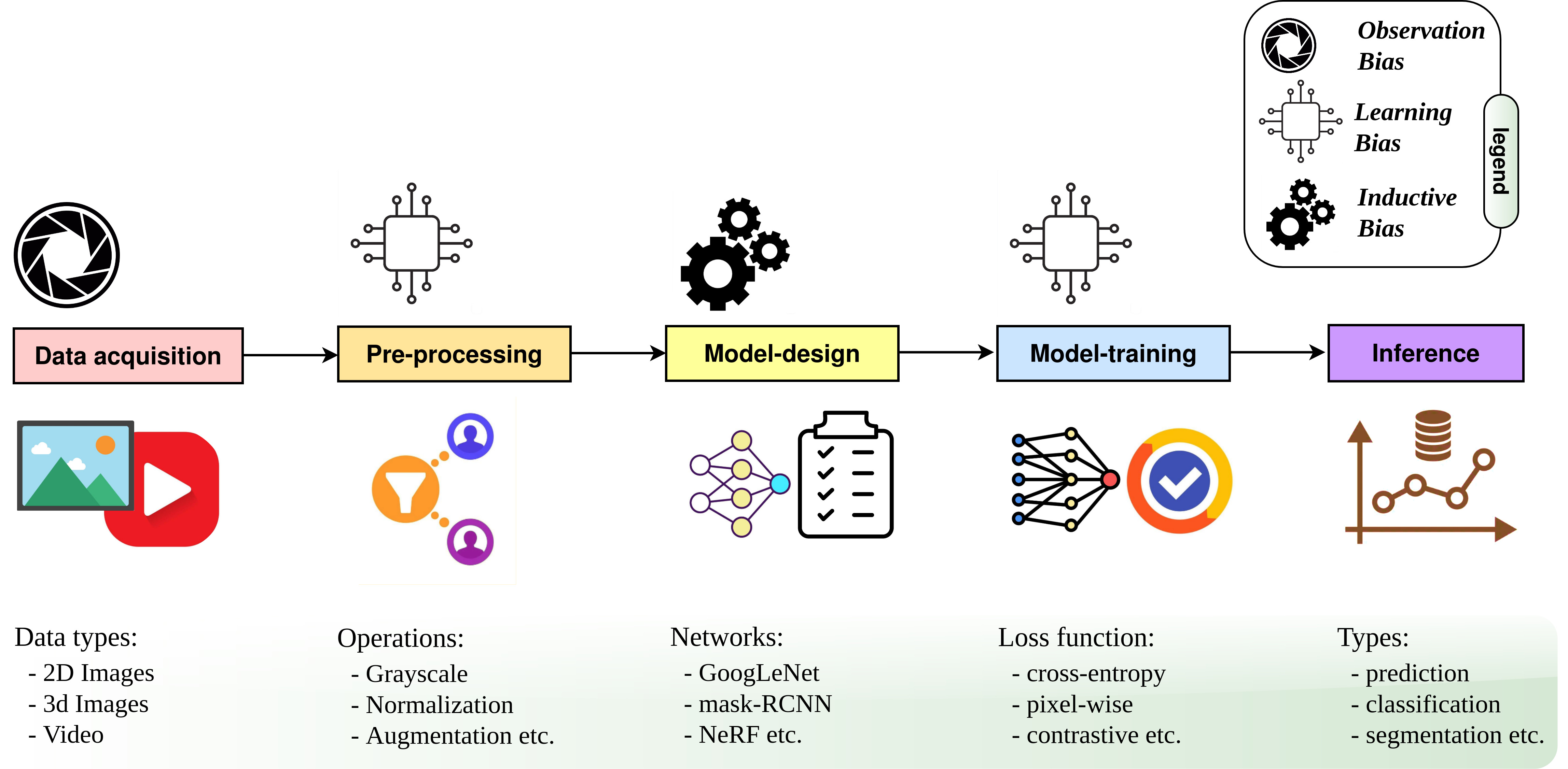}  
  \caption{{\color{black}The computer vision pipeline integrates various physics information through three biases at different stages, beginning with data acquisition (observation bias) from sources like images or videos. Data pre-processing follows, involving cleaning, augmentation, and normalization (learning bias). The model design comes next, selecting architecture elements based on problem requirements (inductive bias). Model training adjusts parameters using pre-processed data, again incorporating learning bias. The pipeline concludes at the inference stage, where the model predicts and evaluates on unseen data.} }
  \Description{}
  \label{fig:CV_pipeline_steps}
\end{figure}

Fig.~\ref{fig:CV_pipeline_steps} integrates a standard CV pipeline with physics information biases to illustrate physics incorporation in PICV, detailed in section~\ref{PIML_intro}. We outline the CV pipeline into five stages: data acquisition, pre-processing, model design, training, and inference, following \cite{elgendy2020deep}, and explore how physics priors are integrated at each stage of the pipeline, with examples in Fig.~\ref{fig:CVpipe_phy_incorp_example}.
%
%
%
Below we provide brief introductions on each of these stages of the CV pipeline and also present an overview of how physics is incorporated into this typical CV workflow. 
\begin{enumerate}
    \item \textbf{Data acquisition:} In this stage, the visual data is input to the computer vision algorithm. The visual data is generally in the form of 2D/ 3D images, videos, and data from specialized sensors (e.g. point cloud data from LIDAR). Physics incorporation at this stage of the CV pipeline falls under the observation bias category (see Fig~\ref{fig:CV_pipeline_steps}). This category is characterized by direct, simulation, or extracted physics data being fed to the computer vision models. For example, in the work by \cite{monakhova2019learned} concerned with lensless imaging, the acquired lensless measurements are fed into a CNN-based custom network which also incorporates the physics of the imaging system, using its point spread function (PSF) see Fig.~\ref{fig:CVpipe_phy_incorp_example}a.

    \item \textbf{Pre-processing:} Acquired visual data is generally non-uniform e.g. different resolutions, color spectrum, etc. as they come from different sources. As a result, each image/ video frame goes through a process of standardization or cleaning up process to make the data ready for the computer vision model. Pre-processing makes the data easy to analyze and process computationally, which in turn improves accuracy and efficiency. Color to grayscale conversion, image standardization and data augmentation (e.g. de-colorize, edge enhancement, and flip/rotate) are some examples of basic pre-processing operations.
    Super-resolution and image synthesis are two popular pre-processing tasks that have been enhanced by PI guidance \cite{kelshaw2022physics,arora2022spatio,
    chen2021physics,siddani2021machine}. For an example see Fig~\ref{fig:CVpipe_phy_incorp_example}b, which relates to the generation CV task.
    The physics incorporation strategies at this stage heavily follow the learning bias approach, characterized by the enforcement of prior knowledge/ physics information through soft penalty constraints. 

    \item \textbf{Model-design:} This phase involves feature extraction and choosing or adapting the model architecture. Techniques like Convolutional Neural Networks (CNN)\cite{lecun1995convolutional,mallat2016understanding}, Graphical Neural Networks (GNN)\cite{bronstein2017geometric}, and others \cite{cohen2019gauge, owhadi2017multigrid, hamzi2021learning, owhadi2019kernel} are used. CNN enhancements for handling symmetries improve applications in fields like medical imaging \cite{winkens2018improved} and climate analysis \cite{cohen2019gauge}. Custom NN models incorporate physical principles directly into their structure, aiding in generalization by enforcing "hard" constraints. Advancements include temporally coherent GANs for fluid dynamics \cite{xie2018tempogan} and CNNs for predicting sea surface temperatures \cite{de2019deep}. For physics-integrated computer vision (PICV), specific network designs, like the P2PSF net for transient image analysis shown in Fig.~\ref{fig:CVpipe_phy_incorp_example}c, extract relevant features. Besides tailored CNN model that integrates physical insights for computer vision tasks \cite{wu2018seeing} depicted in Fig.~\ref{fig:CVpipe_phy_incorp_example}d, aims at computational efficiency and better performance through physics integration.
    
    \item \textbf{Model-training:} CV model training optimizes network parameters through iterative loss minimization, directly influencing model efficiency with functions like cross-entropy or pixel-wise loss. Physics priors, often as PDEs/ODEs, are integrated into this process via the loss function, enhancing learning bias. PI adjustments to the loss function, such as regularization parameters or physics-based loss components, play a crucial role. For instance, in \cite{kissas2020machine}, a PINN architecture parametrizes cardiovascular fluid dynamics solutions with neural networks, training to align with system measurements while adhering to physical laws like the Navier Stokes equation. This introduces physics-based loss components, including momentum and mass conservation, and arterial boundary conditions, showcasing the method's applicability, see Fig.~\ref{fig:CVpipe_phy_incorp_example}e.


\item \textbf{Inference:} This stage in the CV pipeline is concerned with the deployment of the trained models for the prediction of outcomes from new observations. 
There is no physics information induction at this stage since it typically represents the finished/ trained product of the corresponding CV tasks. An as illustrative example Fig.~\ref{fig:CVpipe_phy_incorp_example}f, presents an end-to-end pipeline of a CV prediction task, involving robot motion planning in various cluttered 3D environments, with path solution as the inferred result. The framework represents a wave propagation model generating continuous arrival time to find path solutions informed by a nonlinear first-order PDE (Eikonal Equation). 
\end{enumerate}

\vspace{0.2cm}

%
{\color{black} Integrating physics into the CV pipeline boosts model robustness and accuracy by applying physical principles via soft and hard constraints. Soft constraints, such as PI loss functions, guide models toward physically plausible solutions \cite{kissas2020machine}, while hard constraints embed these principles directly into the architecture, ensuring adherence to physical laws \cite{lecun1995convolutional, bronstein2017geometric, cohen2019gauge}. This blend of approaches, from regularization techniques to custom architectures designed for physical integration \cite{mallat2016understanding, owhadi2017multigrid, hamzi2021learning, owhadi2019kernel, xie2018tempogan, de2019deep, wu2018seeing}, ensures that CV models are both data-driven and fundamentally aligned with the physical world, significantly improving data quality, generalizability, and the physical plausibility of predictions.}

\subsection{Applications of PICV}

This section discusses, in brief, the applications of PICV models in different domains. We have already illustrated the distribution of published papers across application domains in Fig.~\ref{fig:PICV timeline and application statistics}(b). In the following section, we review these application domains in more detail.

\noindent \textbf{Computational imaging and photonics:}
In studies on lensless imaging, custom networks enhance performance \cite{monakhova2019learned}, and PI-based techniques enable video denoising in low light \cite{monakhova2022dancing}. Approaches for DNN generalization in such imagers are explored in \cite{deng2020interplay}. Further, PI methods are advancing imaging across various fields, including through scattering media \cite{zhu2021imaging}, in near-field \cite{chen2022physics} and fluorescence microscopy \cite{xypakis2022physics}, and in elasticity imaging \cite{zhang2020physics}.\newline
\textbf{Robotics:} Recent PI approaches deal with motion planning for robotic agents in cluttered scenarios\cite{ni2022ntfields} and motion synthesis without using motion capture data \cite{xie2021physics}.\newline
\textbf{Surveillance:}
Research in this domain involves intelligent analysis of surveillance videos/ images, with techniques like action recognition \cite{murray2017bio}, pose estimation \cite{xie2021physics}, motion capture \cite{huang2022neural}, tracking \cite{livne2018walking} and crowd analysis \cite{behera2021pidlnet}. \newline
\textbf{Remote Sensing:} 
With regard to urban surface temperature estimation, \cite{wu2022intracity} introduces a PI-estimator for accurate surface temperature prediction, while \cite{chen2022deepurbandownscale} proposed a PI-based network that provides improved high resolution and high precision urban surface temperature downscaling.  
Works like \cite{zantedeschi2020towards,thoreau2022p,fablet2021end} improve prediction and extrapolation capabilities of remote sensing models with variation-prone data.
\cite{lutjens2020physics} provides better present and future high-resolution flood visualization from cloud-obscured images and \cite{liu2022image} generates and auto-annotates hyperspectral images.\newline
\textbf{Weather modeling:} 
Physics-based data-driven approaches are introduced by \cite{chen2021physics} and \cite{yasuda2022super} for troposphere temperature prediction and facilitating real-time high-resolution prediction respectively. 
Papers like \cite{zhang2021three,zhang2021spatiotemporal} proposed a physics-inspired approach for 3-D spatiotemporal wind field reconstruction and spatiotemporal wind field based on sparse LIDAR measurements respectively. In another work, \cite{jenkins2020physics} presented a PI-detection and segmentation approach for gaining insights from solar radio spectrograms.\newline
\textbf{Medicine and Medical imaging:} 
PI approaches have been presented for improved MRI reconstruction \cite{qian2022physics}, conjoined acquisition and reconstruction \cite{weiss2019pilot}, mitigation of imprecise segmentation in differently sourced MRI scans \cite{borges2019physics}, better MRI-based blood flow model \cite{van2022physics}, estimating physiological parameters from sparse MRI data\cite{zapf2022investigating}, cardiovascular flow modeling using 4D flow MRI \cite{kissas2020machine} and for reconstructing single energy CT from dual-energy CT scans \cite{poirot2019physics}.
In heart-function imaging, \cite{buoso2021personalising} simulates left ventricular (LV) bio-mechanics,\cite{sahli2020physics} introduces a PI-network for cardiac activation mapping and \cite{herrero2022ep} simulates accurate action
potential and estimates electrophysiological (EP)  parameter. 
In brain related technologies, \cite{altaheri2022physics} uses encephalogram (EEG) towards motor imagery classification,  \cite{sarabian2022physics} augments sparse clinical measurements and \cite{guc2021fault} performs automatic actuator sensor fault diagnosis, in health monitoring.\newline
\textbf{Geoscience:} 
PI-based approaches of permeability prediction from $\mu$CT scans \cite{garttner2021estimating} and images \cite{wu2018seeing} were proposed. \cite{yang2021revisit}, estimates physically consistent subsurface models using seismograms from geophysical imaging.\newline
\textbf{Dynamical systems:} 
In \cite{kelshaw2022physics} proposed a PI-based approach for super-resolution of sparse, spatial observations of chaotic-turbulent flows. \cite{arora2022spatio}, presents a PI- deep learning-based SR framework to enhance the spatio-temporal resolution of the solution of time-dependent PDEs in elastodynamics. 
\cite{mehta2020physics} introduced a PI-spatiotemporal model to alleviate efficient emulation of crack propagation in brittle materials.\newline
\textbf{Fluid and solid mechanics:} {\color{black}High-fidelity simulations in fluid dynamics and solid mechanics are often prohibitively expensive, leading researchers to use deep learning to enhance data from more computationally manageable coarse-grained simulations. Traditional data-driven methods, though beneficial, generally fail to include essential physical constraints. In contrast, PI methods such as those proposed by \cite{shu2023physics} and \cite{ren2022physics} employ deep neural networks (DNNs) for data reconstruction and achieving super-resolution from low-fidelity inputs, which are critical for enhancing the practical usability of fluid mechanics simulations.

Specific applications in this domain include enhancing the resolution of multiphase fluid flow data \cite{li2022using}, improving spatial resolution of flow fields \cite{gao2021super,fathi2020super,zayats2022super,eivazi2022physics}, and enriching turbulence estimation frameworks \cite{bode2021using,subramaniam2020turbulence}. Recent advancements also focus on reconstructing dense velocity fields from sparse data \cite{wang2022dense,wang2020physics}, estimating various fluid dynamics fields \cite{molnar2023estimating}, generating detailed velocity and pressure fields \cite{siddani2021machine}, and enhancing geostatistical modeling through semantic inpainting \cite{zheng2020physics}. These interdisciplinary approaches demonstrate the convergence of computer vision and mechanics, illustrating a broad and applicable overlap that enriches both fields.}\newline
\textbf{Manufacturing and Mechanical systems:} 
Manyar et al. \cite{manyar2023physics} addressed the detection of anomalous configurations of sheets in the manufacturing process, \cite{du2021physics} combines PI- machine learning, mechanistic modeling, and experimental data to reduce defects in additive manufacturing (AM) process and  \cite{oikonomou2022physics} introduced a PI Bayesian learning framework for auto-calibration of AM technologies. 
Lai et al. \cite{lai2020full}, proposed structural monitoring and vibration analysis using PI based approach with event cameras.\newline
\textbf{Materials science:} 
Here, works have primarily focussed on prediction tasks, such as material fracture pattern prediction from arbitrary material microstructures \cite{wei2022fracture} and composite strength prediction \cite{zhou2021harnessing} from representative volume element (RVE) images. Zhang et al. \cite{zhang2020physics} used PINNs for recovering unknown distribution of material properties.\newline
\textbf{Accident and conflict resolution:} Approaches in this context, attempt to build a PI safety model for estimating crash risk, leveraging historical trajectory data \cite{yao2023physics} and raw video data \cite{arun2023physics}. Another work \cite{zhao2021physics}  is concerned with conflict resolution in air traffic scenarios by leveraging prior physics knowledge

\section{PICV tasks}
\label{sec:PICVtasks}
This section delves deep into computer vision tasks. Using the computer vision pipeline discussed previously, {\color{black} we categorize tasks into 5 primary groups: 
imaging inverse problems (imaging, super-resolution, reconstruction), 
image generation, 
predictive modeling, 
image analysis (classification, segmentation) and 
human analysis.} 
Many works that have been discussed in this survey have multiple computer vision tasks/operations involved in the process and in such cases, we have based our categorization on the particular vision task that has been augmented by incorporation of physics information.
Below, we will briefly discuss tasks before delving deep into each of them.

\subsection{Physics-informed Imaging Inverse Problems}
{\color{black}Inverse imaging problems involve reconstructing unknown data, like signals or images, from observations made through a non-invertible forward process. This challenge is central to tasks such as deblurring, deconvolution, inpainting, and super-resolution, aiming to recover original data from complex outcomes \cite{ongie2020deep}.}
\subsubsection{Physics-informed Imaging}
Imaging captures the world using various modalities. Cameras for RGB, infrared, hyperspectral, and X-ray cover the electromagnetic spectrum. Medical imaging employs ultrasonic, MRI, PET, and CT scans. Sophisticated computer vision algorithms are required to extract information and enhance these images for human interpretation and decision-making. For the latest trend on the imaging task, refer \cite{esteva2021deep,elyan2022computer}.

Monakhova et al. \cite{monakhova2019learned} developed the Le-ADMM-U network for enhanced and rapid computation in lensless imaging, integrating the imaging model's physics through its PSF and parameters derived from lensless camera data. This network extends the ADMM iterative process \cite{boyd2011distributed} by introducing trainable hyperparameters and a deep denoising component.
Similarly, \cite{yanny2022deep} integrates a system's varying PSF into a deconvolution network for single-shot 3D imaging, using a MultiWeinerNet to approximately reverse spatially varying blur.
In addressing low light video denoising under high gain, \cite{monakhova2022dancing} employs a GAN-adjusted PI noise model for camera noise representation. Initially, a generator network (2D U-Net \cite{ronneberger2015u}) with physics-based parameters is trained to create synthetic noisy images/videos. Subsequently, a video denoising network (FastDVDNet \cite{tassano2020fastdvdnet} with HRNet \cite{sun2019deep} blocks) is trained using both synthetic and real clean images/videos.
In lensless multicore fiber (MCF) endoscopy, \cite{guo2023dynamic} effectively generalizes the imaging across varied fiber configurations by initially computing speckle intensity pattern autocorrelations and then applying a U-Net \cite{ronneberger2015u} for object reconstruction. Similarly, Zhu et al. \cite{zhu2021imaging} use a speckle autocorrelation pre-processing and a U-Net post-processing strategy to address generalization challenges in imaging through diffusers or scattering media across different scenes.
{\color{black}Eichhorn et al. \cite{eichhorn2024physics} presents PHIMO, a cutting-edge PI motion correction technique for MRI that enhances T2* quantification by selectively excluding motion-corrupted k-space lines, thus shortening acquisition times. Zhu et al. \cite{zhu2023physics} introduce a PI Sinogram Completion method for CT imaging that minimizes metal artifacts while avoiding over-smoothing, leveraging physical principles and novel algorithms. Kamali et al. \cite{kamali2023elasticity} employ PINN to accurately determine the elastic properties of materials, showing promise for advanced biomedical imaging. Bian et al. \cite{bian2023high} achieve significant advancements in single-photon imaging quality through deep learning and physical noise modeling. Halder et al. \cite{halder2023mri} develop MRI-MECH, a PINN-based framework to improve the diagnosis of esophageal disorders by modeling fluid dynamics and mechanical health metrics in dynamic MRI. Lastly, Yang et al. \cite{yang2023fwigan} propose FWIGAN, a novel unsupervised framework using GANs for 2D full-waveform inversion in geophysics, surpassing existing FWI methods by addressing initial model sensitivity and data noise issues without the need for labeled data or pretraining.}

A number of papers use PINNs as an efficient approach to introduce physics information in deep learning. 
In optical microscopy, \cite{chen2022physics} employs PINNs grounded in full-vector Maxwell’s equations for the inverse retrieval of photonic nanostructure properties like electric permittivity and magnetic permeability from near-field data. Zhang et al. \cite{zhang2020physics} utilize PINNs for identifying mechanical property distributions in elasticity imaging, incorporating PDEs, boundary conditions, and incompressibility constraints for hyperelastic materials. Similarly, Saba et al. \cite{saba2022physics} apply PINNs to predict scattered fields in diffraction tomography, using the Helmholtz equation as a physical loss.

\begin{figure}[h!]
\centering
\def\arraystretch{1}%
\scalebox{1}{
\begin{tabular}{ll}
\subf{\includegraphics[width=0.38\columnwidth]{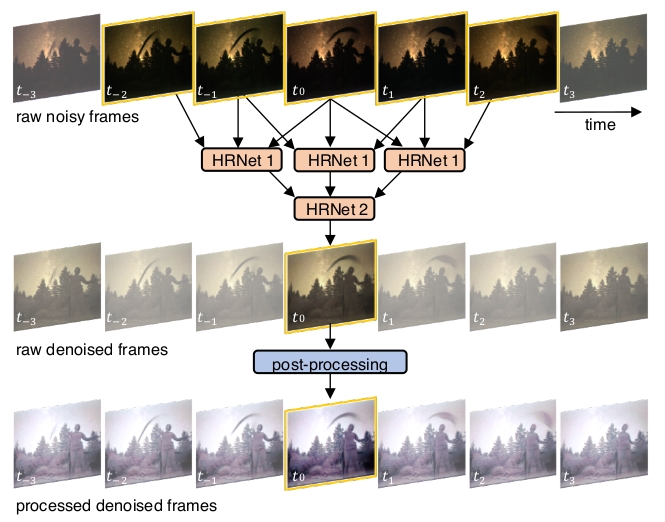}}
     {(a) Imaging \cite{monakhova2022dancing}}
&
\subf{\includegraphics[width=0.5\columnwidth]{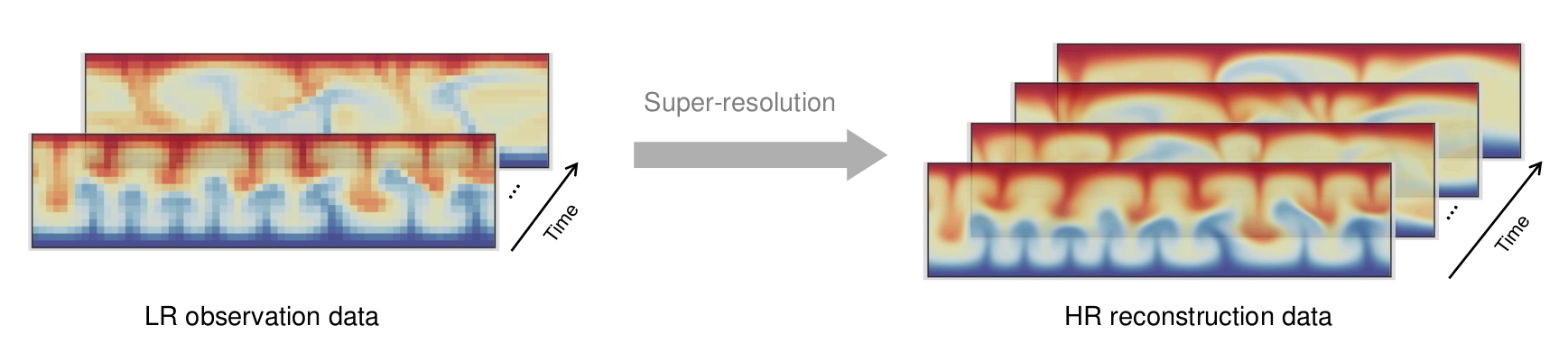}}
     {(b) Super-resolution \cite{ren2022physics}}
\\
\end{tabular}}
\caption{Examples of CV tasks \textbf{(a)} Imaging \cite{monakhova2022dancing}, the shown HRNet denoising network is trained using a GAN tuned physics based camera noise model, for photorealistic low-light video denoising,  \textbf{(b)} Super-resolution \cite{ren2022physics}, the schematic shows low-resolution coarse grid data of a 2d Rayleigh-Benard convection system w.r.t. temperature on the left and its high resolution reconstructed form on the right.}
\Description{}
\label{fig:examples_Imaging_sr}
\end{figure}

\subsubsection{Physics-informed Super-resolution}
Super-resolution aims to generate higher-quality images from low-resolution inputs using trained models, enhancing beyond the original training image quality. It finds key applications in surveillance \cite{aakerberg2022real,jung2021adaptive,farooq2021human} and medical imaging \cite{georgescu2023multimodal,zhang2022soup,gu2020medsrgan}, with detailed reviews available in \cite{wang2020deep,anwar2020deep}. For tackling sparse and noisy data, deep learning proves effective in super-resolution, yet incorporating PI approaches ensures model outputs adhere to physical principles.

Kelshaw et al. \cite{kelshaw2022physics} enhanced sparse chaotic-turbulent flow observations using PI-CNN, integrating physics in the loss term to align high-resolution outputs with underlying PDEs. Arora et al. \cite{arora2022spatio} improved spatial and temporal resolution of coarse PDE solutions using a PI residual dense network (RDN \cite{zhang2018residual}), eliminating the need for high-resolution data. Bode et al. \cite{bode2021using} applied a GAN with super-resolution, adversarial, and PI losses, employing 3D-CNN \cite{krizhevsky2017imagenet} and RRDB for enhanced turbulent flow statistics. Li et al. \cite{li2022using} adapted SRGANs (SRGAN \cite{ledig2017photo}) for multiphase fluid flows by incorporating a physics-based loss in the discriminator, ensuring precise high-resolution turbulent flow reconstructions.
PhySRNet, presented in \cite{arora2022physrnet}, generates deformation fields in hyperelastic materials without requiring high-resolution annotations, using separate networks based on residual dense networks (RDN) \cite{zhang2018residual} to enhance solution fields from low-resolution simulations.

\cite{yasuda2022super} developed the SE-SRCNN model for high-resolution, real-time temperature forecasts in urban areas, leveraging skip connections, channel attention, and specific feature extractors for various inputs like temperature and building height. \cite{ren2022physics} introduced the PhySR network for enhancing spatiotemporal scientific data, combining ConvLSTM networks \cite{shi2015convolutional} for temporal refinement with pixel shuffling for detailed spatial reconstruction. \cite{gao2021super} proposed a CNN-SR model that enhances resolution and determines parameters, using a PDE loss function based on the Navier Stokes theorem for integrating physical principles. Subramanium et al. \cite{subramaniam2020turbulence} presented the PI method with TEGAN (from SRGAN \cite{ledig2017photo}) for improved turbulence representation, employing PI loss functions similar to \cite{raissi2017physics} for physically-consistent enhancements.

Several SR techniques utilize PINNs for enhancement tasks. Eivazi et al.~\cite{eivazi2022physics} use PINNs to refine spatio-temporal flow-field data from sparse and noisy inputs, bypassing the need for high-resolution references. To enhance the spatio-temporal quality of 4D-Flow MRI, PINN-based models are applied in \cite{fathi2020super}, following the incompressible Navier-Stokes (NS) equations and mass conservation principles during network training. Moreover, \cite{zayats2022super} introduces a SPINN method for estimating turbulent flows from low-resolution inputs, relying on NS equations without defined initial conditions or forcing.
\cite{wang2020physics} developed a Physics-Informed Neural Network (PINN) called PINSSR for plume simulation super-resolution, integrating RRDB blocks \cite{wang2018esrgan} with a physics consistency loss. This approach minimizes the physics residual, based on advection-diffusion equations, between high-resolution and improved low-resolution images.

{\color{black} Burns et al. \cite{burns2023untrained} discusses an untrained PI neural network for flexible SIM reconstruction, while \cite{shu2023physics} presents a diffusion model improving CFD data super-resolution through PI conditioning, particularly for 2D turbulent flows. The ETSSR technique \cite{haghighi2023accelerating} accelerates stereo image simulation for autonomous driving with a novel transformer architecture. SRUNK-Res \cite{fok2023deep}, a super-resolution U-Net, and a PINN method for 4D-flow MRI data \cite{shone2023deep} show significant advances in medical imaging by enforcing physics-based constraints. A 3D CNN and GAN-based model \cite{trinh20243d}, a U-net framework for DFM images \cite{lei2024super}, SpkSRNet for optical speckle patterns \cite{li2023learning}, a T2-deblurred SR method for 3D-TSE MRI \cite{chen2023physics}, and a Variational Super-Resolution Neural Network for CFD \cite{pradhan2023variational} all demonstrate substantial improvements in resolution and quality, showcasing robustness across various applications and conditions.}

\subsubsection{Physics-informed Reconstruction}
There are two typical methods for reconstructing images: tomography image reconstruction and image recovery reconstruction. Tomography imaging involves creating a single image of an object that has been imaged in sections. Deep image reconstruction, or deep learning-assisted tomography reconstruction, is widely used in fields such as oceanography, remote sensing, and material science.
For recent trends in deep image reconstruction refer to \cite{wang2020deeptomo}.
The second type focuses on enhancing degraded or incomplete images, known as image recovery. Image recovery and corresponding reconstruction approaches find their applications in computational imaging \cite{batchuluun2020thermal,rivenson2018phase} and in medical imaging \cite{de2019deep,gong2018pet}. For a detailed discussion on image recovery refer \cite{liu2018image}.

Poirot et al. \cite{poirot2019physics} introduced a method for reconstructing non-contrast single-energy CT (SECT) images from dual-energy CT (DECT) scans. This method employs a CNN (inspired by ResNet\cite{he2016deep}) that takes advantage of the physical principles behind DECT image creation and the insights obtained from training on real images, resulting in DECT images of enhanced quality.
Shu et al. \cite{shu2023physics} introduced a method for reconstructing high-fidelity computational fluid dynamics (CFD) data from low-quality inputs using a denoising diffusion probabilistic model (DDPM) \cite{ho2020denoising}. By treating CFD data reconstruction as a denoising task, and incorporating PI conditioning via PDE residual gradients during training and sampling, they significantly enhanced reconstruction accuracy.
\cite{fablet2021end} proposed an end-to-end architecture for the reconstruction and forecasting of sea surface dynamics from irregularly sampled satellite images. The framework consists of a variational model with cost minimization through physics-driven parametrization of the flow operator and also consists of an LSTM-based solver model. 

Zhang et al. \cite{zhang2021three} developed a technique to reconstruct 3-D wind fields over time, merging the dynamics of 3-D Navier-Stokes (NS) equations with LIDAR observations. This approach employs a neural network that integrates LIDAR data and NS equations, guided by a specialized loss function incorporating both NS and LIDAR constraints. {\color{black} The PI Fringe Pattern Analysis (PI-FPA) \cite{yin2024physics} utilizes a streamlined neural network alongside Fourier transform profilometry, enhancing phase retrieval in optical metrology. By incorporating physical principles, it offers improved single-shot 3D imaging across diverse samples.}
In fluorescence microscopy image denoising, \cite{xypakis2022physics} introduces a novel DNN architecture, RESUNET. It integrates physics by employing data normalization with a photon model and a PI loss function reflective of the Poisson-distributed photon detection process.


\begin{figure}[h!]
\centering
\def\arraystretch{2}%
\scalebox{1}{
\begin{tabular}{ll}
\subf{\includegraphics[width=0.42\columnwidth]{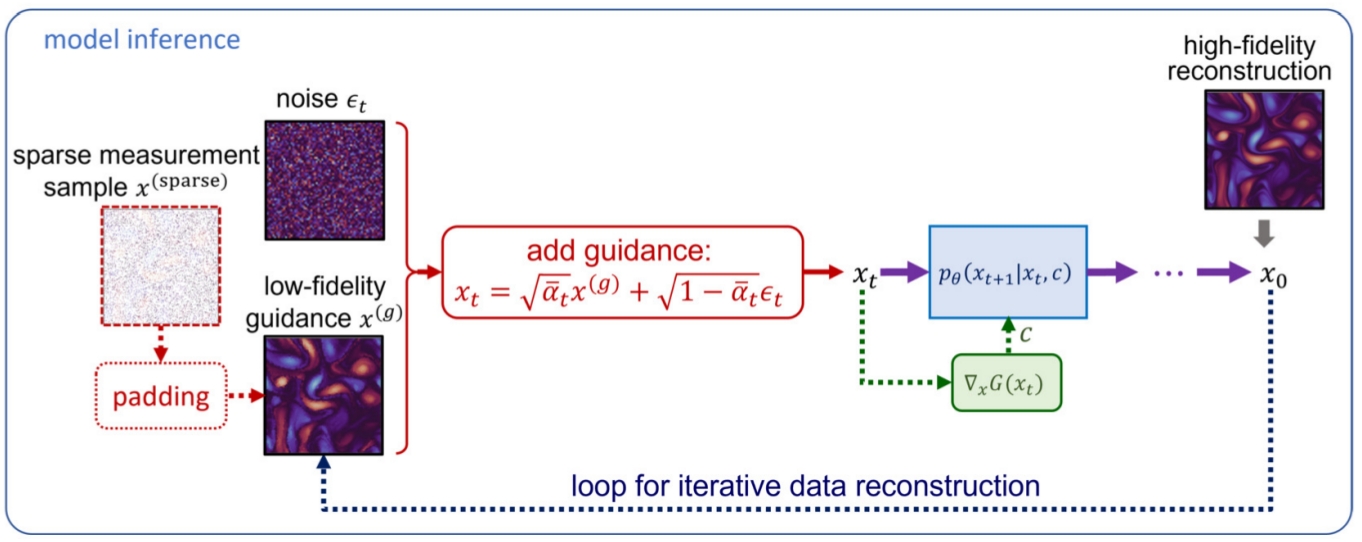}}
     {(a) Reconstruction \cite{shu2023physics}}
&
\subf{\includegraphics[width=0.45\columnwidth]{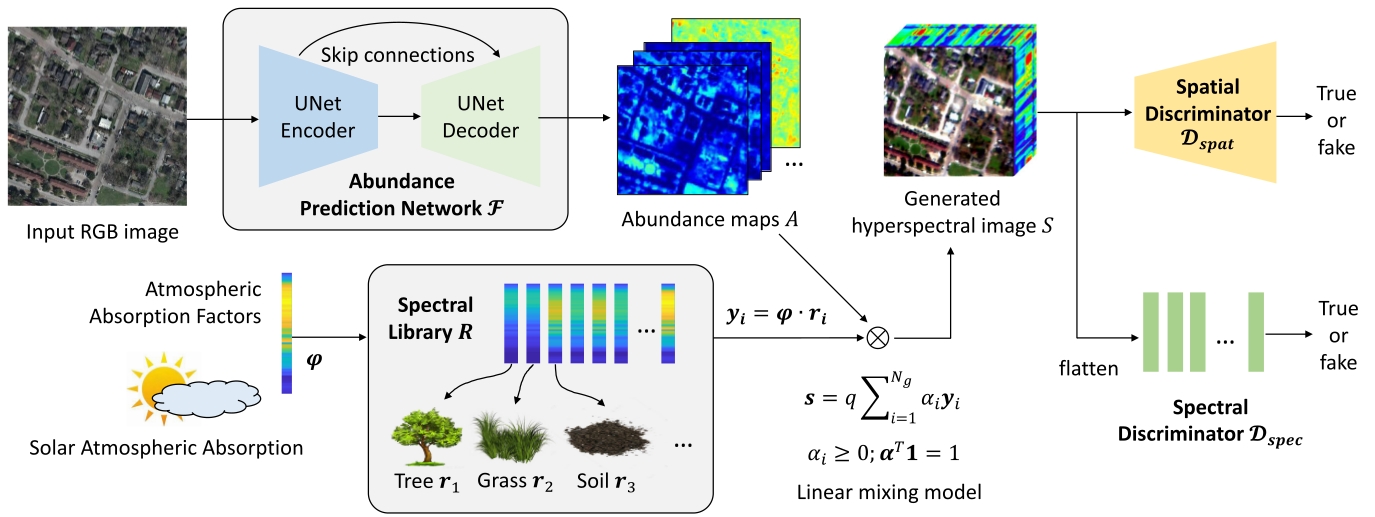}}
     {(b) Generation \cite{liu2022physics}}
\\
\end{tabular}}
\caption{Examples of PICV tasks \textbf{(a)} Reconstruction \cite{shu2023physics}, shows the inference phase of a diffusion model, which reconstructs high-fidelity data from either a low-fidelity sample or a sparsely measured sample, while guided by PI conditioning information, \textbf{(b)} Generation \cite{liu2022physics}, the workflow elaborates the synthesis of high-quality spectral data and generation of subpixel-level spectral abundance, for remote sensing application. } 
\Description{}
\label{fig:examples_recon_gen}
\end{figure}

In \cite{wang2022dense}, a method using Physics-Informed Neural Networks (PINNs) reconstructs dense velocity fields from sparse data acquired by particle image and tracking velocimetry techniques, optimizing a loss function that incorporates both experimental data and the NS equations. Similarly, \cite{molnar2023estimating} presents a PINN approach, PI-background-oriented schlieren (BOS), to derive density, velocity, and pressure fields from reference and distorted image pairs in fluid dynamics. This method leverages a physics loss derived from Euler and irrotationality equations, ensuring the flow fields comply with both experimental observations and fundamental physics. 
To simulate personalized left ventricular (LV) biomechanics, \cite{buoso2021personalising} introduced a PINN-based method. This approach allows for patient-specific customization, creating functional cardiac models from clinical images with minimal computational expense. It relies on a shape model (SM) derived from high-resolution cardiac images for approximating LV anatomies and a function model (FM) based on LV anatomies' displacement fields calculated via a biophysical finite element (FE) model. The FM underpins the physics-based final layer of the PINN.
Qian et al. \cite{qian2022physics} introduced a physics-informed deep diffusion-weighted MRI reconstruction technique called PIDD. This method synthesizes multi-shot DWI data using a physics-based motion phase model and then employs a deep learning network trained on this synthetic data for robust, high-quality image reconstruction. 

Neural Radiance Field (NeRF) methods model scenes by rendering multi-view images with neural networks under ground-truth supervision \cite{chen2022aug}. However, NeRF's novel view interpolations often result in visually inconsistent and geometrically rough outputs. Recent studies aim to bridge this 'generalization gap' for unseen views by integrating physical principles.
Chu et al. \cite{chu2022physics} developed a technique for reconstructing flow motion in hybrid scenes containing fluids and obstacles, without needing initial conditions or information on boundaries and lighting. This method combines image data, physical principles, and a GAN-based data prior model \cite{chu2020learning} within a PINN framework. Similarly, Li et al. \cite{lipac} introduced PAC-NeRF for system identification in the absence of geometric priors, blending physical simulations with rendering. This enables the estimation of both geometric and physical properties from multi-view videos, enhancing NeRF with a continuum dynamics model based on the material point method (MPM) \cite{jiang2015affine}. \cite{chen2022aug} introduces Aug-NeRF, a three-tier augmented NeRF training pipeline with physical grounding. It enhances geometry reconstruction, generalizes better for unseen views, and shows robustness against noisy inputs by incorporating a prior across coordinates, MLP intermediate features, and pre-rendering MLP outputs with distinct physical meanings.
{\color{black} \cite{mondal2024physics} introduces a deep learning-based, PI model to simulate underwater image effects, addressing color distortion and low contrast due to light attenuation and scattering. The work uses a complex image formation model to generate realistic ground truth images and hard-coding a basic image formation equation into the network, which infers additional factors affecting underwater image degradation. \cite{ning2023image} presents a novel VPIN method for enhancing optical synthetic aperture images beyond traditional MAP frameworks by integrating variational inference with deep learning and employing Residual Dense Blocks for superior feature extraction. \cite{delcey2023physics} showcases the first use of PINNs for 3D reconstruction of unsteady gravity currents from sparse data, leveraging LAT-2PIV—a combination of light attenuation technique and particle image velocimetry—to embed physical equations into PINNs through automatic differentiation, achieving high accuracy and cost efficiency in capturing complex, transient hydrodynamic fields.
}

\subsection{Physics informed Image Generation}
Image generation presents a significant challenge in computer vision due to data's high dimensionality. Generative models, essential for tasks like image editing \cite{kim2021exploiting,zhu2020domain}, fusion \cite{ma2020pan,xu2020mef}, synthesis \cite{brock2018large,zhan2019spatial}, domain adaptation \cite{huang2018auggan,zhang2022exposing}, and data augmentation \cite{frid2018gan,mariani2018bagan}, have gained traction. Notably, Generative Adversarial Networks (GANs) have advanced, producing realistic images within specific constraints \cite{sorin2020creating,alqahtani2021applications}. Recent reviews detail GANs' roles in medical image generation \cite{singh2021medical}, augmentation \cite{chen2022generative}, and remote sensing applications \cite{jozdani2022review}.

\cite{chen2021physics} presents a PI-generative neural network (PGnet) for troposphere temperature prediction, utilizing a two-stage approach that includes PI propagation based on the DCNet architecture \cite{yang2021deconvolution} and physics-agnostic generation, leveraging convection-diffusion PDE constraints. 
\cite{lutjens2020physics} employs a modified pix2pixHD\cite{wang2018high} network, incorporating physics through a flood extent map, to produce satellite images depicting coastal flooding scenarios before and after the event, achieving physics-based image-to-image translation.
The PDASS method \cite{liu2022physics} generates high-resolution hyperspectral images and precise subpixel annotations from an RGB image using a U-Net\cite{ronneberger2015u}-based adversarial training framework that incorporates physics insights like imaging mechanisms and spectral mixing.
Siddani et al. \cite{siddani2021machine} developed a GAN-based method that, once trained, generates synthetic velocity and pressure fields around randomly distributed particles, accounting for non-dimensional variables, local coordinates, and discrete symmetries to include physics. Meanwhile, \cite{zheng2020physics} enhances semantic inpainting for geostatistical modeling by incorporating physical data through direct and indirect measurements, utilizing the Wasserstein Generative Adversarial Network with Gradient Penalty (WGAN-GP) \cite{gulrajani2017improved}.
{\color{black}Yang et al. \cite{yang2023physics} discusses the development of IPADS within DL for MRI, highlighting its role in creating synthetic data, minimizing the need for real data, and improving scalability, explainability, and privacy in biomedical imaging. Kawahara et al. \cite{kawahara2023mri} presents a framework for synthesizing FLAIR and DWI images from T1- and T2-weighted MRIs, using conditional adversarial networks to enhance the versatility and quality of multi-contrast MRI synthesis without extra scans. Pan et al. \cite{pan20232d} employs a diffusion model with a Swin-transformer network, effectively generating synthetic medical images to overcome dataset limitations and significantly improve the diversity and quality of synthetic images for AI training.}

In PhysDiff \cite{yuan2022physdiff}, the authors integrate denoising diffusion (DDPM) \cite{ho2020denoising} and physical constraints into a diffusion model for human motion modeling. They introduce a physics-based motion projection module for motion imitation in simulations, ensuring physical constraints are met.
\cite{manyar2023physics} generates PI, photo-realistic synthetic images to identify anomalies and defects in composite layup processes. This is achieved by training a mask region-based convolutional network (Mask-RCNN) \cite{he2017mask} with both real and synthetic images. The paper employs a physics-based simulator for creating synthetic images of sheet defects. However, image-to-image (i2i) translation, used for transferring images between domains, faces quality issues due to the entanglement effect. 
Pizzati et al. \cite{Pizzati2023PIGAN}  presents a disentanglement method, where they use a collection of simple physical models rendering some of the physical traits (e.g. water drop, fog, etc.) of these phenomenons and learn the remaining ones. 

%
\renewcommand{\arraystretch}{1.1}
\begin{table*}[htpb]
\centering
\caption{Characteristics of PICV literature w.r.t. different computer vision tasks.} \label{table: PICV characteristics} 
\resizebox{\textwidth}{!}{%
\begin{tabular}{|l|l|l|l|l|l|l|}
\hline
\multirow{9}{*}{\begin{turn}{270}\textbf{Imaging}\end{turn}} &\textbf{Ref.} &\textbf{Context} & \textbf{Physics guided operation} & \textbf{Training dataset} & \textbf{DNN/CV Model } &  \textbf{Physics information}\\[0.5ex] 
\hline
\hline
& &  & &  &  &\\
&\cite{monakhova2022dancing} &
Low light imaging & Video denoising & Custom & GAN  & Noise model parameters\\
&\cite{monakhova2019learned} & Lenseless imaging & Image computation & Custom & Le-ADMM-U & PSF \\
&\cite{deng2020interplay} & Lenseless imaging  & Cross dataset generalisation & ImageNet \cite{deng2009imagenet}, Face-LFW\cite{huang2008labeled} & Customised PhENN \cite{sinha2017lensless}  & Training dataset \\
& & && IC-layout \cite{goy2018low} and MNIST\cite{lecun1998mnist} & &\\
&\cite{yanny2022deep} & 3D imaging & Image sharpening  & Custom & MultiWeinerNet & PSF\\
&\cite{guo2023dynamic} & Endoscope imaging  & Image computation & MNIST \cite{lecun1998mnist,xiao2017fashion} & CNN & Speckle auto-correlation \\
&\cite{zhu2021imaging} & Scattering imaging & Generalised & MNIST \cite{lecun1998mnist}, & CNN & Speckle\\
& & & image reconstruction & FEI face \cite{thomaz2012fei} & & correlation \\
&\cite{chen2022physics} & Near-field microscopy & Parameter retrieval & Custom & PINN & PDE (Maxwell's equation)\\
&\cite{zhang2020physics} & Elasticity imaging & Material identification & Custom & PINN & PDE, BC hyperelastic material\\
&\cite{poirot2019physics} & Medical imaging (CT) & High fidelity CT processing & Custom & Custom (based on  & Lookup virtual non-\\ 
& & & & & ResNet \cite{he2016deep}) & contrast (L-VNC) image\\
&\cite{weiss2019pilot} & Medical imaging (MRI) & Accelerated MRI & NYU fastMRI initiative database \cite{zbontar2018fastmri}, & Custom  & Physical MRI hardware \\
& & & & Medical segmentation decathlon \cite{simpson2019large} & &constraints (e.g. slew rate)\\
&\cite{xypakis2022physics} & Fluorescence
microscopy & Image denoising & Custom & RESUNET\cite{zhang2018road}  & Physical loss\\
&\cite{saba2022physics} & Diffraction tomography & Tomograhic reconstruction & Custom & PINN & PDE (Maxwell's equation)\\
&\cite{eichhorn2024physics}& Medical imaging (MRI) & Motion correction &- & Custom & signal evolution physics \\
&\cite{zhu2023physics}& Artifact removal & Sinogram completion & Custom &- & Beam hardening \\
&&&&&& Correction Model\\
&\cite{kamali2023elasticity}& Biomedical imaging & Identify tumor edges & Custom & PINN & Linear elastic theory,\\
&&&&&&  physical measurement\\
&\cite{bian2023high} & High speed imaging & single photon imaging & SPAD & Swin-Transformer based & Physical noise model\\
&\cite{halder2023mri}& Medical analysis & Esophageal disorder diagnosis & custom & PINN & Fluid flow eqns.,\\
&&&&&&conservation laws\\
&\cite{yang2023fwigan}&Geophysics & subsurface imaging & Marmousi(1,2)& GAN based on WGAN & Acoustic wave equation\\
&&&&&&\\
\hline

\multirow{12}{*}{\begin{turn}{270}\textbf{Super-resolution (SR)}\end{turn}} & & & & & & \\
&\cite{yasuda2022super} & Micro-meteorology & Estimates temp. fields & Custom & Custom SRCNN & Sim. data (LES)\\
&\cite{kelshaw2022physics} & Dynamical system  & SR of chaotic flow  & Custom  & VDSR 
& PDE \\
&\cite{arora2022spatio} & Dynamical system  & Spatio-temporal SR & Custom  & Custom (based on \cite{zhang2018residual})& IC, BC, PDE\\
&\cite{eivazi2022physics} & Fluid mechanics & SR based data augmentation & \cite{borrelli2022predicting}  &PINN  & PDE (Burgers eqn.)\\
&\cite{ren2022physics} & Dynamical systems  & Spatio-temporal SR & Simulated using \cite{esmaeilzadeh2020meshfreeflownet}  & ConvLSTM 
& PDE, BC (Dirichlet, Neumann\\
&\cite{fathi2020super} & 4D Flow MRI & SR and denoising & custom, CFD simulated & PI-DNN  & PDE(NS), mass conservation\\
&\cite{gao2021super} & Fluid flow  & SR and denoising & CFD sim using\cite{jasak2007openfoam} & PI-CNN & PDE (NS) loss, BC\\
&\cite{zayats2022super} & 2D turbulent flow  & zero shot SR & generated using NSE & custom PI-CNN  & Luenberger observer \\
&\cite{bode2021using} & Turbulent  & sub-filter modeling & Decaying  & custom (based on  & Physical loss term\\
& & reactive flows & & turbulence DNS \cite{gauding2019self} & ESRGAN 
& \\
& \cite{li2022using} & Multi-phase  & SR & Generated using \cite{jasak2007openfoam} & Custom (based on & Algebraic loss term\\
& & fluid simulation & & \say{DamBreak} case & SRGAN 
& (Interphase equations)\\
&\cite{wang2020physics} & Atmospheric pollution & SR in advection & Simulated (using & Custom (based on  & sim. training data, physics- \\
&&plume  model  & diffusion models  & adv.-diff. eqn.)& ESRGAN 
& consistency loss\\
&\cite{arora2022physrnet} & Solid mechanics  & SR of deformation fields & Generated using \cite{alnaes2015fenics}  & Custom (based on \cite{zhang2018residual}) & PDE, Constitutive law\\
&\cite{subramaniam2020turbulence} & Turbulence enrichment & Generation & CFD simulation & Custom (based on -  & sim. data and physics - \\
&&  & &  & SRGAN 
& loss (continuity, pressure)\\
& \cite{shu2023physics} & Fluid dynamics  & CFD construction & - & DDPM (PINN) & PDE and conditioning loss  \\
&\cite{haghighi2023accelerating} & Driving simulation & Stero image simulation  & CARLA sim. & Custom ETSSR  & Disparity maps \\
&\cite{fok2023deep} & medical imaging  & X-ray SR & Custom & SR UNet (SRUNK)  &  modulation TF kernel \\
&\cite{shone2023deep} & 4D Flow MRI  & SR & CFD & PINN & NS eqns.\\ 
&\cite{trinh20243d} & Aerodynamics  & Custom CFD generated & CFD & based on 3D CNNs and GANs  & Poisson and Continuity eqns.  \\
&\cite{li2023learning} & Imaging & SR & custom & Modified ResNeXt-101  & Implicit physical \\
& &  &  & - &  & char. of speckle patterns  \\
&\cite{chen2023physics} & Biomedical imaging (MRI) & Scan time reduction  & GEMM MR image & SRGAN & Physical variables \\
&\cite{pradhan2023variational} & Fluid dynamics  & Multiscale modeling & Custom & custom VSRNN & VMS formulation  \\
& &  &  & - &  &  \\

\hline

\multirow{9}{*}{\begin{turn}{270}\textbf{Reconstruction}\end{turn}} &&&&&&\\ 
&\cite{shu2023physics} & Fluid mechanics  & high-fidelity computational fluid  & 2-D Kolmogorov flow & Denoising Diffusion- & PDE residual gradient\\
& &  & dynamics simulation data reconstr. &  & Probabilistic Model (DDPM) &\\
&\cite{zhang2021three} & Fluid dynamics  & Wind field reconstruction & Custom (LIDAR measurement) & Custom (PINN based)  & PDE (3D NS equations)\\
&\cite{wang2022dense} & Flow visualization  & Velocity reconstruction & DNS dataset & PINN  & PDE (NS equations)\\
&\cite{molnar2023estimating} & Supersonic flow  & Field and parameter estimation & Custom  & PINN  & PDE (Euler, irrotationality eqns.)\\
&\cite{chen2022aug} & Physical simulation  & Image augmentation, denoising & LLFF, NeRF-Synthetic & MLP & Worst case perturbations \\
&\cite{chu2022physics} & Fluid dynamics & Smoke reconstruction & ScalarFlow dataset & Custom & NS equations\\
&\cite{ning2023image} & Space telescopy & aperture imaging & Golay-6 & VPIN & OSA degradation model\\
&\cite{delcey2023physics} & Geophysics & Flow analysis & custom & PINN & Flow equations\\
&\cite{mondal2024physics} & Image restoration & Underwater image correction & Custom & Custom (DenseNet-169)  & image dehazing model\\
&\cite{yin2024physics} & Optical meterology &  Fringe pattern analysis & - & Custom Lightweight DNN & Fourier Transform Profilometry\\
&\cite{burns2023untrained} & Microscopy & SR image reconstr. & BioSR & PINN & Illumination process model \\
& &  & &  &  &\\
\hline

\multirow{6}{*}{\begin{turn}{270}\textbf{Image generation}\end{turn}} & & & & & & \\
&\cite{chen2021physics} & Troposphere  & Temperature field & ERA5 \cite{hersbach2020era5} & Custom  & Physical process data \\
&& temperature prediction  & generation & &  &(motion field), mask loss\\
&\cite{siddani2021machine} & Fluid flow  & Generate and pressure fields & DNS sim. results \cite{moore2019hybrid}  & Custom GAN & Sim. training data \\
&\cite{zheng2020physics} &  Geostatistical modeling & Semantic inpainting & Generated using \cite{harbaugh2005modflow}& WGAN-GP\cite{gulrajani2017improved}  & PDE, physical constraints\\
&\cite{lutjens2020physics} & Flood visualization & Pre and post flood & xBD \cite{gupta2019xbd}  & pix2pixHD \cite{wang2018high}& flood map, evaluation metric\\
& &  & image generation & Flood maps (SLOSH-NOAA) &  &\\
& \cite{liu2022physics}& Hyperspectral image  & Generation & USGS Spectral Library\cite{kokaly2017usgs} &   & Abundance map, spectral library\\
&& synthesis & & IEEE $grss\_ dfc\_ 2018$ , GF5 datasets \cite{liu2021hyperspectral} & &\\
&&&&&&\\
\hline
\end{tabular}}
\label{table:PIML_CV_char_imaging}
\end{table*}

%
\renewcommand{\arraystretch}{1.1}
\begin{table*}[htpb]\ContinuedFloat
\centering
\caption{(Contd.) Summary of PICV literature w.r.t. different computer vision tasks} 
\resizebox{\textwidth}{!}{%
\begin{tabular}{|l|l|l|l|l|l|l|}
\hline
%
\multirow{16}{*}{\begin{turn}{270}\textbf{Image generation (contd.)}\end{turn}} &\textbf{Ref.} &\textbf{Context} & \textbf{Physics guided operation} & \textbf{Training dataset} & \textbf{DNN/CV Model } &  \textbf{Physics information}\\
\hline

& \cite{thoreau2022p} & Semantic segmentation & Generative model & Simulated using DART\cite{gastellu2012dart} & Custom (based on & Latent physical variables \\
&&  & &  & $\phi$-VAE \cite{takeishi2021physics})  & \\
&\cite{qian2022physics} & Imaging  & Image synthesis  & Synthesized via multi-shot & Custom  & Polynomial motion phase model\\
& &  & for reconstruction &  DWI data synthesis  &  & \\
&\cite{manyar2023physics} & Defect detection & Image generation  & Custom (Real+ Sim.) & ResNet-50  & Simulated input data\\
&\cite{oviedo2019fast} & X-ray classification & Data augmentation & Custom  & CNN  & Domain knowledge (particularities -\\
&&&&&& of thin-film XRD spectra)\\
&\cite{Pizzati2023PIGAN}& Robotics/ & I2I translation & Custom & Custom & Physical model\\
&& autonomous driving & feature disentanglement &&&\\
&\cite{yang2023physics}&Biomedical MRI & synthetic data generation & - & Custom & Bloch equations\\
&\cite{kawahara2023mri}& Medical imaging & generation &Custom& GAN& MR properties\\
&\cite{pan20232d}& Medical imaging & data augmentation & ACDC MRI, BTCV etc. & Custom (MT-DDPM) & diffusion process\\
&\cite{borges2024acquisition} & Medical MRI analysis & Data augmentation & custom, ABIDE  &  UNet &\\
&\cite{lei2024super} & Dark-field microscopy (DFM) & data augmentation & Simulated MATLAB2020 & UNet based CNN  & DFM physics model \\
& & & & & &\\
%

\multirow{16}{*}{\begin{turn}{270}\textbf{Predictive modeling}\end{turn}} & & & & & &\\
\hline
& & & & & &\\
&\cite{arun2023physics} & Traffic safety  & Safety field model learning & Custom  & - & Model parameters\\
&\cite{yao2023physics} & Accident prevention & Vehicle safety prediction & HIGH-SIM\cite{shi2021video} & Custom (CNN-LSTM) & Historical trajectory data\\
&\cite{zhang2021spatiotemporal} & Weather  & Wind-field prediction  & Custom (LoS wind speed values)  & Custom PINN  & Loss terms (NS, LIDAR measure.)\\
&\cite{ni2022ntfields} & Robot navigation & Motion planning & Computed using a Speed Model  & Custom & PDE, Collision-avoidance constraint\\
&\cite{mehta2020physics} & Dynamical systems & Coupled-dynamics emulator & Custom  & Custom (based on-   & Spatiotemporal derivatives,\\
&&&&& ST-LSTM 
& Loss function, Sim. data\\
&\cite{garttner2021estimating} & Hydrodynamics  & Permeability estimation & Segmented X-ray $\mu$CT scans 
& Custom (CNN based)& Physics input (Max. flow value)\\
&\cite{zapf2022investigating} & Medicine & Diffusion coefficient estimation & Custom & PINN  & 4D PDE\\
&\cite{herrero2022ep} & Medicine  & Electrophysiological & Simulated cardiac EP data  & PINN  &   PDE, ODE, IC and BC\\
&&  & parameter estimation  & using FD solver &&\\
&\cite{kissas2020machine} & Cardiovascular flow & Predicting arterial- & Synthesized using DG solver 
& PINN  & Conservation law constraints\\
&& modeling & blood pressure &  &  &  (mass, momentum)\\
&\cite{zhao2023physics}& Thermal analysis & Thermal field prediction &  \cite{chen2021deep}& Custom UNET(PINN) & Heat conduction equations\\
&\cite{sarabian2022physics} & Medicine & Brain heamodynamics & Custom & PINN & 1D ROM PDE, Constraints (conser-\\
&&  & prediction &  &  & vation of mass, momentum)\\
&\cite{van2022physics} & Myocardial perfusion (MP) & MP MRI quantification  & Custom  & PINN & ODE residual loss\\
&\cite{sahli2020physics} & Cardiac electrophysiology  &Cardiac activation mapping & Custom & PINN  & PDE (Eikonal equation)\\
&\cite{oikonomou2022physics} & Manufacturing  & Learning Jet printing dynamics & S1, S2 from \cite{hrynevich2020accurate} & - & ODE, BC\\
&\cite{zhou2021harnessing} & Materials & Composite strength prediction & Custom & Custom CNN, VGG16 \cite{simonyan2014very}  & RVE patterns\\
&\cite{wei2022fracture} & Materials & Fracture pattern prediction & Generated using LPM  & Customised FCN\cite{long2015fully} & Sim data (LPM), NN phy. constraint\\
&\cite{chen2022deepurbandownscale} & Weather  & Surface temperature estimation & LST, NDVI, Atmosphereic & Custom PINN  & Multimodal high-resolution data\\
&&  & & forcing, 3D point cloud &  & \\
&\cite{zantedeschi2020towards} & Weather  & Precipitation forecasting & SimSat, ERA5, IMERG   & Custom  & Reanalysis dataset ERA5 \cite{hersbach2020era5}\\
&\cite{zhao2021physics} & Conflict resolution & RL Policy learning & Simulated & CNN & SSD based image\\
&\cite{fablet2021end} & Satellite altimetry & Prediction of Sea & Based on the NEMO model, & RESNET & Multimodal data\\
&&& surface dynamics & NATL60 configuration &&\\
&\cite{buoso2021personalising} & Biophysical modeling & Cardiac mechanics simulation &  MMWHS \cite{buoso2019reduced} & PINN  & NN projection layer,\\
&&  & &  &  & cost function\\
&\cite{wu2018seeing} & Materials  & Fast permeability prediction & Custom/ generated & PI-CNN  & Physical data inputs\\
&&  & &  &  & (porosity, surface area ratio)\\
&\cite{lipac} & Geometry agnostic & Physical parameter estimation &  Custom & MLP & Conservation law,\\
&& System identification &&&& Eulerian-Lagrangian representation\\
&\cite{muller2022deep}& Geophysics & Velocity model building & Custom & U-Net & Surrogate velocity model\\
&\cite{yang2021revisit}& Geophysics & Seismic waveform inversion & Custom & Custom WGAN & 2D acoustic wave eqn.\\
&\cite{cai2021flow} & Fluid dynamics &  Estimate
velocity, pressure fields & Tomo-BOS  & PINN & PDE(NS equations)\\
&\cite{ma2022risp} & Robotics & Identify params. from video & Custom & RISP (custom) & Dynamics model\\
&\cite{lopez2023warppinn} & Medical/Cardiac health & Cardiac strain estimation & Custom SSFP-MRI  & PINN  & Near-incompressibility of cardiac tissue \\
& & & & & &\\
\hline

\multirow{6}{*}{\begin{turn}{270}\textbf{Class.}\end{turn}} &&&&&&\\ 

&\cite{guc2021fault} & Health monitoring & Fault cause assignment & Custom & DCNN, GoogLeNet 
& Time-frequency representations\\
&\cite{altaheri2022physics} & Brain computer interface & Motor imagery classification & BCI-2a dataset & Custom & EEG input data\\
&\cite{du2021physics} & Materials  &  Defect prediction & Custom & - & Mechanistic variables\\
&\cite{lai2020full} & Vision based monitoring  & Structural vibration tracking & Custom  & -  & Basis function\\
& &  & and analysis &  &  & for boundary condition of beams\\
&&&&&&\\
\hline

\hline

\multirow{4}{*}{\begin{turn}{270}\textbf{Seg.}\end{turn}}& & & & & & \\
&\cite{jenkins2020physics} & Solar radiography  & Segmentation of solar radio-bursts & Custom & - & Solar burst drift model\\
&\cite{borges2019physics} & Brain imaging & Brain MRI segmentation & Custom, SABRE subsets & 3D U-Net
& Physics parameter as training input\\
& &  & &  &  &\\

\hline

\multirow{11}{*}{\begin{turn}{270}\textbf{Human analysis}\end{turn}} & & & & & & \\
&\cite{murray2017bio} & Action recognition (AR)  & AR model learning & JHUMMA dataset & HMM  & Acoustics from micro-doppler sensor\\
%
%
&\cite{gartner2022differentiable} & 3D motion reconstruction & Pose estimation & Human3.6M, AIST & Custom & Physics constraints and simulator\\
&\cite{gartner2022trajectory} & 3D pose reconstruction & Pose estimation & Human3.6M, HumanEva-I, AIST & HUND+SO+GT+Dynamics models & Physics engine\\
&\cite{isogawa2020optical} & 3D pose estimation  & Estimate 3D pose sequences & Custom  & Custom  & Physics simulator\\
&\cite{yuan2021simpoe} & 3D pose estimation  & Pose estimation from monocular video &  Human3.6M, Custom & Custom  & Physics simulator\\
&\cite{xie2021physics} & Motion estimation, synthesis  & Motion synthesis model & Human3.6 M & Custom & Physics loss\\
&\cite{zhang2022pimnet} & Motion estimation & Prediction model  & H3.6M MoCap dataset & LSTM Encoder-decoder arch. & Physics dynamics model\\
&\cite{huang2022neural} & Motion capture  & Distribution prior training & Human3.6 M, GPA, 3DOH, GPA-IM & Custom & Human-scene interaction, human -\\
&&  & &  &  & shape reference, physics simulator\\
&\cite{yuan2022physdiff} & Motion generation   & Motion diffusion model & HumanAct12, UESTC & Custom & Physics simulator\\
&\cite{livne2018walking} & Motion capture & Motion tracking & Hasler dataset & - & Physical constrains\\
%
&\cite{behera2021pidlnet} & Video analysis & Crowd characterization & Kinetics dataset & Custom  & Physical parameters \\
&&  & &  &  &(entropy and order)\\
& \cite{chen2021grounding} & Video analysis  & Dynamic video reasoning & CLEVRER \cite{Yi*2020CLEVRER:}  & Custom & Observation and language context\\
&\cite{xu2023interdiff} & Human-object interaction (HOI)  & predicting 3D HOI & BEHAVE & custom DDPM & PI interaction predictor NW\\
& \cite{yi2022physical} & Motion capture  & Motion optimization & Custom  & Custom & Physics simulator\\
& &  & &  &  &\\



\hline
\hline
\end{tabular}}
\label{table:PIML_CV_char_imaging}
\end{table*}

\subsection{Physics Informed Image analysis}
\subsubsection{Physics-informed Image Segmentation}
Image segmentation divides an image into segments or subgroups to simplify analysis by reducing complexity. This process involves labeling pixels according to their categories. Segmentation techniques are categorized into instance, semantic, and panoptic segmentation — the latter being a blend of the first two. Instance segmentation \cite{gu2022review} identifies, segments, and classifies individual objects within an image, organizing pixels around object boundaries and distinguishing between overlapping objects without prior knowledge of their class.
Semantic segmentation \cite{mo2022review} categorizes each pixel into distinct classes without considering additional context. Panoptic segmentation \cite{li2022survey} merges semantic and instance segmentation, distinguishing and identifying each object instance, thus yielding highly detailed information. These segmentation techniques are widely applied in fields such as medical imaging \cite{hatamizadeh2022unetr,tang2022unified}, robotics \cite{narasimhan2022self,jokic2022semantic}, and autonomous driving \cite{wang2022sfnet,fantauzzo2022feddrive}.

In \cite{jenkins2020physics}, the authors introduce a method for detecting and segmenting type II solar bursts in solar radio spectrograms, integrating a drift model of signal frequencies to enhance detection accuracy and training efficiency. An adaptive region of interest (ROI) technique is also proposed to focus on areas matching the burst curvature at specific frequencies, utilizing HOG features and logistic regression without neural networks. \cite{weiss2019pilot} presents PILOT (PI Learned Optimized Trajectory), combining physics insights and deep learning to optimize MRI scan acquisition and reconstruction, accelerating the process. This approach uses physical constraints on magnetic gradients to inform the design of acquisition trajectories, implemented in a unified network that performs both tasks, concluding with a U-Net \cite{ronneberger2015u}-based model for final image reconstruction or segmentation.

In \cite{borges2019physics}, the authors tackle the challenge of inaccurate MRI image segmentation by combining a convolutional neural network with multi-parametric MRI simulations. This approach, integrating MR sequence parameters into a 3D U-Net \cite{cciccek20163d}, enhances the network's robustness to MRI variations across different sites. Meanwhile, \cite{thoreau2022p} introduces a PIML model, P$^{3}$VAE, for semantic segmentation in high-resolution hyperspectral images, aiming for better extrapolation and interpretability. This model evolves from physics-integrated VAEs \cite{takeishi2021physics}. {\color{black} Borges et al. \cite{borges2024acquisition} introduces an algorithm that boosts MRI segmentation by mitigating site biases using PI data augmentation, uncertainty assessment, and harmonization, enhancing cross-site applicability and reducing the need for manual annotations. Similarly, WarpPINN, a PI neural network, improves cardiac strain estimates from cine MRI through precise deformation tracking, applying tissue incompressibility, and utilizing Fourier feature mappings, showcasing its efficacy in cardiac imaging \cite{lopez2023warppinn}.}

\subsubsection{Physics-informed Image Classification} 
Image classification assigns labels to image pixels or vectors based on rules derived from spectral or textural features, using unsupervised \cite{richards2022clustering}, supervised \cite{richards2022supervised}, semi-supervised \cite{liu2020semi,yalniz2019billion}, or self-supervised \cite{azizi2021big,wang2021transpath} methods. Unsupervised classification clusters data without training samples, while supervised classification utilizes labeled training data and methods like \say{maximum likelihood} and \say{minimum distance} to categorize images \cite{shinozuka2009synthetic}. Semi- and self-supervised techniques are also effective in various applications.

\begin{figure}[h!]
\centering
\def\arraystretch{1}%
\scalebox{1}{
\begin{tabular}{ll}
\subf{\includegraphics[width=0.5\columnwidth]{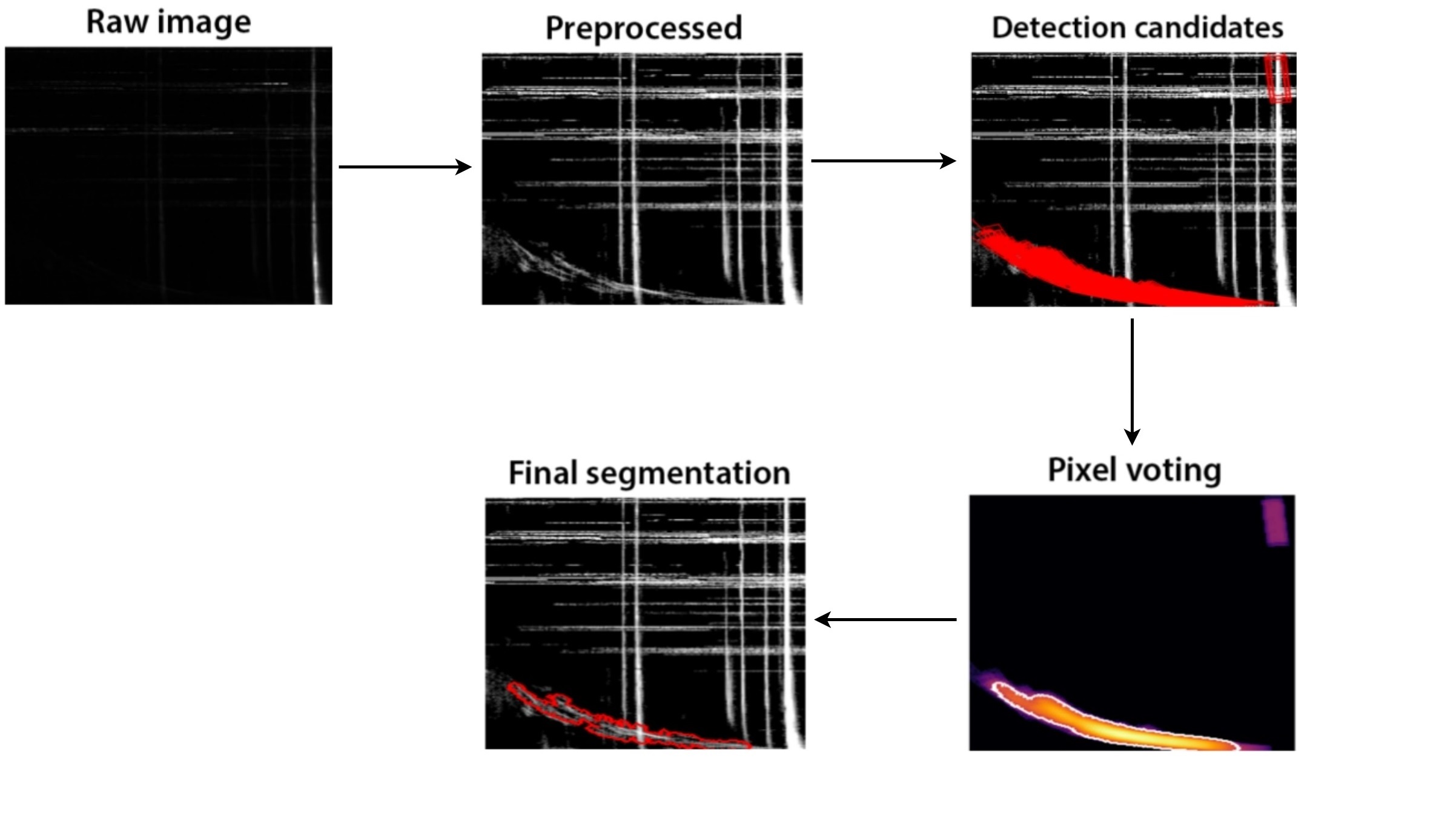}}
     {(a) Segmentation \cite{jenkins2020physics}}
&
\subf{\includegraphics[width=0.36\columnwidth]{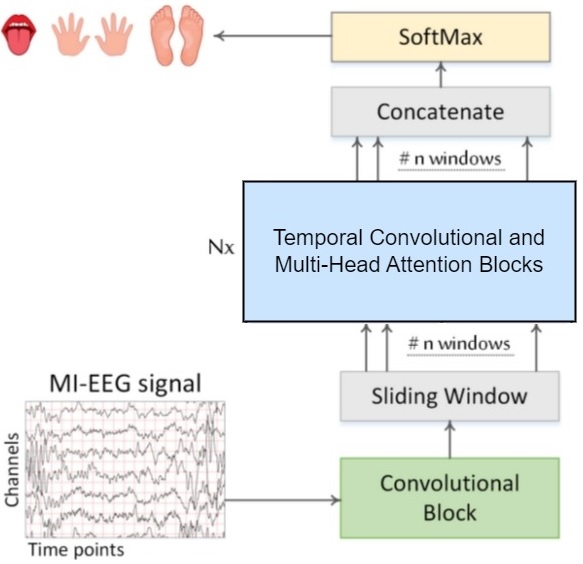}}
     {(b) Classification \cite{altaheri2022physics}}
\\
\end{tabular}}
\caption{Examples PICV tasks \textbf{(a)} Segmentation \cite{jenkins2020physics}, shows the stages of detection and segmentation of occurrence of type II solar bursts in solar radio-spectrograms. The prior knowledge of how such bursts drift through frequencies over time is crucial for the method, \textbf{(b)} Classification \cite{altaheri2022physics}, shown here is the workflow of EEG-based motor imagery (MI) classification algorithm, which uses a novel attention-based temporal convolutional network for boosting classification performance. } 
\Description{}
\label{fig:examples_segment_class}
\end{figure}

Guc et al. \cite{guc2021fault} introduced a method for automatic sensor and actuator fault diagnosis using dynamic mode decomposition with control (DMDc) \cite{proctor2016dynamic} and input-output data. DMDc, by leveraging system measurements and external controls, unveils the system dynamics. These dynamics are then analyzed in time-frequency domains using transfer learning with GoogLeNet DCNN \cite{szegedy2015going} to identify various sensor bias faults.

The ATCNet, introduced in \cite{altaheri2022physics} for EEG-based motor imagery (MI) classification, comprises three key components: a CV block that transforms raw MI-EEG signals into compact sequences, a multihead self-AT block emphasizing crucial information within these sequences, and a temporal convolution (TC) block for extracting advanced temporal features. The CV block modifies EEGNet \cite{lawhern2018eegnet} by using 2-D instead of separable convolution. The AT block incorporates a multihead self-attention layer \cite{vaswani2017attention}, and the TC block employs the TCN framework \cite{ingolfsson2020eeg}. 
Du et al. \cite{du2021physics} integrate PI-machine learning with mechanistic modeling and experimental data to mitigate defects in additive manufacturing. They identify crucial variables from defect formation data, revealing underlying physics, and compute these through a mechanistic model for use in their methodology.

\subsection{Physics-informed Predictive Modeling}
Predictive tasks in computer vision, such as forecasting events or labeling, utilize deep learning models trained on visual datasets. These models, enhanced by computer vision techniques, efficiently extract vital features from images for more accurate predictions. For instance, the paper \cite{domico2022machine} applies image processing and unsupervised learning to identify sunspot features for geomagnetic storm forecasting, correlating them with the \say{Kp-index} through supervised learning. For further information, see \cite{cheng2021fashion, arshad2019computer}.
Yao et al. in \cite{yao2023physics} introduce a Physics-Incorporated real-time Safety Prediction (PMSP) model for vehicle safety, utilizing historical trajectory data to enhance deep network training for predicting vehicle safety indicators. Similarly, Zhao et al. in \cite{zhao2021physics} employ a PI reinforcement learning approach for air traffic conflict resolution, using a Solution Space Diagram (SSD) that integrates key flight data as a physics prior, with a CNN-based RL framework to derive optimal conflict resolution policies.

\cite{wu2022intracity} developed a PI-hierarchical perception (PIHP) network for predicting future urban surface temperatures with high precision and resolution, utilizing multispectral satellite imagery and informed by physical processes for enhanced accuracy in LST forecasting. \cite{garttner2021estimating} employs a PI CNN (PhyCNN) to estimate permeability from micro-CT scans of geological samples, combining direct numerical simulation results and physical characteristics, including flow, porosity, and surface area, for refined predictions.
To enhance precipitation forecasting from satellite data, a three-phase approach involving state estimation, forecasting, and precipitation prediction was developed by \cite{zantedeschi2020towards}. This process integrates physics knowledge using the ERA5 reanalysis dataset \cite{hersbach2020era5} in training a convolutional LSTM model \cite{shi2015convolutional}, enabling it to mimic atmospheric dynamics. For composite strength prediction, Zhou et al. \cite{zhou2021harnessing} employed a custom CNN and a VGG16 \cite{simonyan2014very} network for transfer learning. They used sampled images of representative volume elements (RVE) analyzed with the finite element method for composite damage, training deep learning models to predict composite strength from RVE images directly.

\begin{figure}[h!]
\centering
\def\arraystretch{2}%
\scalebox{1}{
\begin{tabular}{ll}
\subf{\includegraphics[width=0.38\columnwidth]{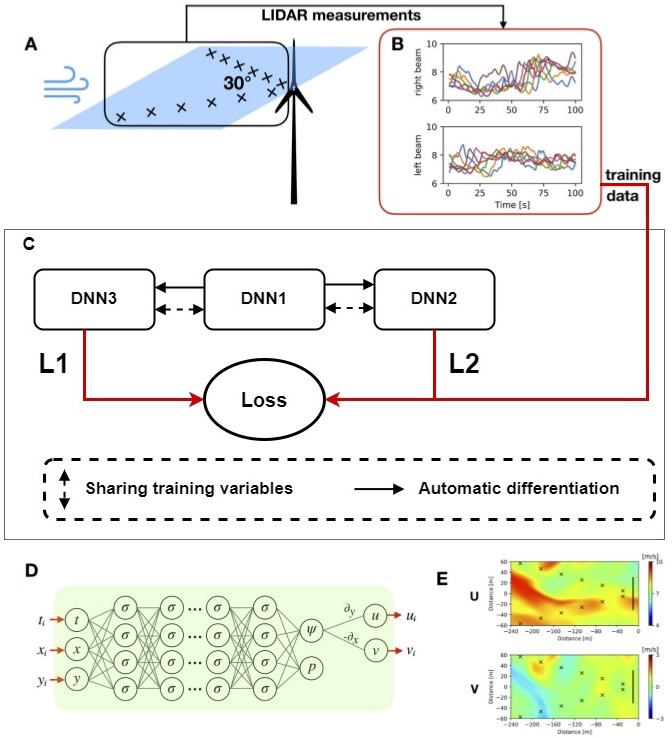}}
     {(a) Predictive modeling \cite{zhang2021spatiotemporal}}
&
\subf{\includegraphics[width=0.5\columnwidth]{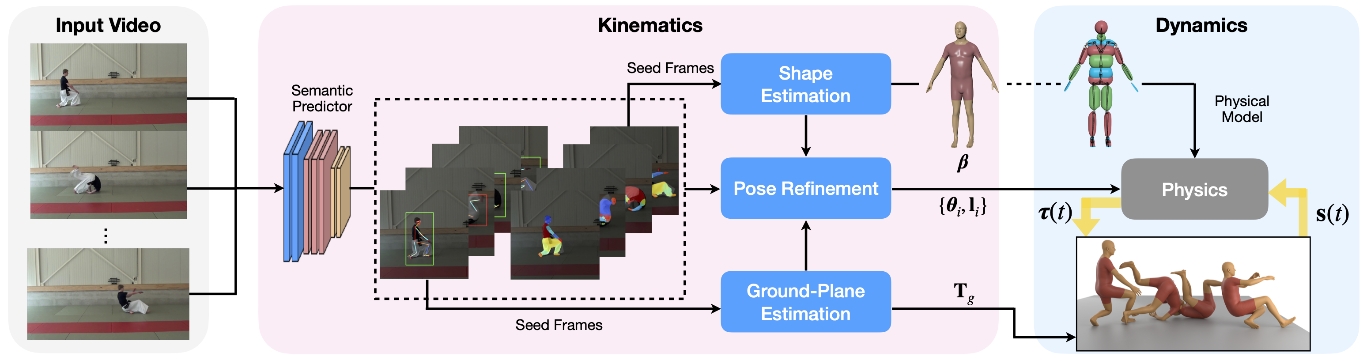}}
     {(b) Human analysis task, human pose estimation \cite{gartner2022trajectory}}
\\
\end{tabular}}
\caption{Examples PICV tasks \textbf{(a)} Predictive modeling \cite{zhang2021spatiotemporal}, the figure presents the workflow of a spatiotemporal wind field prediction method, which works by combining LIDAR measurements and flow physics information, \textbf{(b)} Human pose \cite{gartner2022trajectory}, this overview shows that, with input of an unknown real-world video, the algorithm estimates the ground plane location and dimensions of the physical body model. It then recovers the physical body motion aided by a fully featured physics engine in this human pose estimation process.} 
\label{fig:examples_PredModel_Human_analy}
\end{figure}

Deng et al. \cite{deng2020interplay} demonstrate how public datasets like ImageNet\cite{krizhevsky2017imagenet} improve DNN generalization in lensless imaging, outperforming low-entropy datasets like MNIST\cite{lecun1998mnist} using PhENN\cite{sinha2017lensless}. Ni et al. \cite{ni2022ntfields} introduce a PI-motion planner that uses a wave-based model for navigation in clutter, combining various neural network components. Mehta et al. \cite{mehta2020physics} develop a model simulating material stress over time, integrating physics into its computations. 

A DNN for predicting fracture patterns using the lattice particle method is presented in \cite{wei2022fracture}, incorporating physical constraints and microstructure analysis. Chen et al. \cite{chen2022deepurbandownscale} create the DUD framework using PINNs for more accurate urban temperature forecasts, while Zapf et al. \cite{zapf2022investigating} apply PINNs to extract physiological measures from MRI scans. A method for inferring physical system dynamics from video under varying conditions is introduced in \cite{ma2022risp}, and {\color{black} a PI-CNN that predicts temperature fields without labeled data is detailed in \cite{zhao2023physics}.}
EP-PINN, which simulates action potentials and estimates EP parameters from limited data, is discussed in \cite{herrero2022ep}. Kissas et al. \cite{kissas2020machine} use PINNs for cardiovascular flow modeling from non-invasive MRI data. Sarabian et al. \cite{sarabian2022physics} and \cite{sahli2020physics} apply PINNs to brain hemodynamics and cardiac activation mapping, respectively. Wu et al. \cite{wu2018seeing} present a PI-CNN for porous media analysis.

In geophysics, \cite{yang2021revisit} proposes a physics-informed approach for seismic FWI using unlabeled data, and \cite{muller2022deep} integrates supervised learning with PINNs to improve FWI. A PINN framework for myocardial perfusion MR is outlined in \cite{van2022physics}, and \cite{arun2023physics} introduces a model for crash risk assessment from video data. Zhang et al. \cite{zhang2021spatiotemporal} and Cai et al. \cite{cai2021flow} apply PINNs to wind and flow analysis, respectively.
Lai et al. \cite{lai2020full} and \cite{oikonomou2022physics} demonstrate the effectiveness of PI models for structural monitoring and electro-hydrodynamic additive manufacturing calibration, emphasizing the integration of physical principles and machine learning.

\subsection{Physics-informed Human Analysis Tasks}

\subsubsection{Human analysis} Yuan et al. introduced SimPoE, a method combining kinematic inference and physics-based control for 3D human pose estimation, using 3D scene modeling for physical contact integration \cite{yuan2021simpoe}. Gartner et al. enhanced 3D pose estimation with a physics model for plausible motion generation from video, using DiffPhy for motion refinement \cite{gartner2022trajectory}\cite{gartner2022differentiable}. InterDiff forecasts 3D human-object interactions with a focus on full-body dynamics and physical accuracy by merging diffusion model predictions with PI refinements \cite{xu2023interdiff}. Xie et al. employed a PI optimization for motion synthesis from video poses, refining motions for improved pose estimation and future motion prediction using a time-series model and a smooth contact loss function \cite{xie2021physics}\cite{scheithauerjorge}\cite{yuan2020dlow}. Isogawa et al. presented a method for 3D pose extraction from photon histograms, enhancing accuracy with a learnable inverse PSF function \cite{isogawa2020optical}.

For human motion capture, Huang et al. and Zhang et al. introduced models that generate physically plausible motions by incorporating real physical supervision, human-scene interactions, and physical constraints into the training and denoising processes \cite{huang2022neural}\cite{zhang2022pimnet}. Murray et al. utilized micro-Doppler sonar with RGB-depth data for recognizing human actions through HMM, linking Doppler modulations to pose sequences \cite{murray2017bio}. Behera et al. introduced PIDLNet, employing physics-based features for analyzing structured versus unstructured crowd movements \cite{behera2021pidlnet}. Livne et al. offered a generative 3D human pose tracking method, integrating physical constraints without needing prior scene or subject knowledge \cite{livne2018walking}. 
{\color{black}\cite{yi2022physical} introduces the Physical Inertial Poser (PIP), a cutting-edge six-sensor motion capture method, overcoming the drawbacks of traditional techniques with its neural-physical integration, enhancing motion capture's accuracy and realism.

\subsubsection{Analysis and scene understanding}
Significant progress has been made in tracking and scene understanding. Ma et al.\cite{ma2023learning} developed the NCLaw framework that integrates neural networks with PDEs to improve motion data learning through differentiable simulation. Deng et al.\cite{deng2023learning} introduced the DVP method for predicting fluid dynamics from videos, increasing both accuracy and physical realism. Further advancements by VRDP~\cite{ding2021dynamic} and HyFluid\cite{yu2024inferring} combine visual, linguistic, and physics-based techniques to enhance dynamic predictions and reconstruct fluid dynamics from sparse data. Additionally, the Dynamic Concept Learner (DCL)~\cite{chen2021grounding} advances dynamic visual reasoning by focusing on object tracking and interaction, predicting video outcomes with minimal supervision using both linguistic and visual cues.
}

{\color{black} 
\section{Quantitative study and insights}
\label{sec: quant_analysis}
 \subsection{Quantitative Analysis of PI Enhancements in CV Tasks}
In this discussion we highlights representative examples from select CV tasks, illustrating the significant impact of PI.
In \textit{\textbf{Inverse Imaging}}, Guo et al. \cite{guo2023dynamic} demonstrated robust generalization in imaging with perturbed fibers, achieving PSNRs of 44.78 dB for seen objects and 22.83 dB for unseen objects. Monakhova et al. \cite{monakhova2022dancing} enhanced video denoising in low-light, achieving a PSNR of 27.7 and SSIM of 0.931, surpassing FastDVDnet by 16.4\% in PSNR and 50.6\% in SSIM.
For \textit{\textbf{Generation}}, PI GANs in flood visualization \cite{lutjens2020physics} achieved an IoU of 0.553 and a lower LPIPS of 0.263. Zheng et al. \cite{zheng2020physics} demonstrated superior semantic inpainting with PI constraints, achieving RMSE of ~0.02 and SSIM over 0.98.
In \textit{\textbf{Predictive Modeling}}, Sahli et al. \cite{sahli2020physics} improved median RMSE in activation times for atrial fibrillation from 3.92 ms to 1.53 ms in homogeneous and from 4.77 ms to 2.23 ms in heterogeneous scenarios. The EP-PINNs model \cite{herrero2022ep} delivered RMSEs as low as 6.0 $\times$ $10^{-3} \pm 2.0 \times 10^{-3}$, even under noisy conditions where RMSEs could reach up to 9.0 $\times$ $10^{-3} \pm 4.0 \times 10^{-3}$, in context of arrhythmia treatment.
For \textit{\textbf{Classification}} tasks, \cite{guc2021fault} addressed system faults with 98.7\% accuracy, while the PI ATCNet \cite{altaheri2022physics} achieved 85.38\% accuracy and a kappa score of 0.81, marking a 4.71\% improvement.
In \textit{\textbf{Segmentation}}, \cite{jenkins2020physics} adopted an adaptive curved ROI for solar bursts segmentation, enhancing the Global IoU from 0.134-0.146 to 0.229-0.312. In MRI, \cite{borges2019physics} used a PI based segmentation approach that raised Dice scores for Grey Matter from 0.904 to 0.910 and for White Matter from 0.943 to 0.948, with significant p-value (<0.0001) improvements. 
In \textit{\textbf{Human synthesis}}, \cite{isogawa2020optical} showed that a PI approach reduced MPJPE from 123.9 to 96.1 in single-subject evaluations and from 137.7 to 108.6 cross-subject. \cite{xie2021physics} noted a PI approach decreased MPJPE by 7.5\%, Global Root Position Error by 42.6\%, esmooth Error by 26.2\%, Foot Tangential Velocity Error by 41.7\%, and Foot Height Error by 80.3\%, improving pose accuracy and realism.
}

{\color{black}
\subsection{When to choose PICV over typical data-driven CV approaches ?}
In specific scenarios, a PICV-based approach can offer more dependable, precise, and adaptable solutions than purely data-driven CV approaches. 
For instance, when data is sparse or limited, such as in remote sensing, PICV can enhance image interpretation and fill data gaps, leading to more precise analysis. For complex physical phenomena, like those encountered in medical imaging (MRI, CT), PICV leverages deep knowledge of tissue properties and imaging techniques to improve reconstructions and anomaly detection. In applications where accuracy and reliability are critical, such as autonomous vehicle navigation, incorporating models of light reflection and motion dynamics can significantly improve object detection and scene understanding, particularly under challenging conditions like severe weather. PICV is also beneficial in industries like semiconductor manufacturing or aerospace, where data acquisition costs are high, by enhancing defect detection through models of material properties and manufacturing processes. Additionally, in environmental modeling and prediction, PICV excels by integrating multi-modal data and applying physical laws to forecast changes accurately, such as in flood or forest fire predictions.

}

{\color{black}
\subsection{Cross-Domain Synergies and Innovations}
The adoption of physics-informed (PI) methodologies demonstrates a significant trend in combining deep physical principles with data-driven models across various fields, leading to the creation of robust, accurate, and efficient models. These models are not only domain-specific but also share insights and methods across different areas. In sectors like computational imaging, photonics, and remote sensing, PI methods have enhanced image reconstruction and denoising \cite{monakhova2019learned, chen2022deepurbandownscale}, similar advancements seen in MRI technologies \cite{qian2022physics}. These improvements reflect the use of physical models to better interpret complex data, extending to weather modeling \cite{chen2021physics} and remote sensing \cite{wu2022intracity} for improved analysis and prediction. In robotics, motion planning \cite{ni2022ntfields}, and surveillance, PI models facilitate understanding dynamic environments, akin to their role in fluid and solid mechanics \cite{shu2023physics} for simulation accuracy and data enhancement. Materials science also benefits from PI in predicting properties and behaviors \cite{wei2022fracture}, with similar applications in geology and manufacturing \cite{garttner2021estimating, manyar2023physics}, highlighting the cross-disciplinary utility of physics-based modeling. Additionally, PI methodologies extend to dynamical systems \cite{kelshaw2022physics} and resolving accidents \cite{yao2023physics}, showcasing the broad applicability of physics in addressing complex, evolving challenges and fostering innovation across diverse domains.

\section{Open-questions and gaps in research}
\label{sec:openresearch}
\subsection{Open questions in PICV}
In this section, we discuss in brief the crucial challenges in the extensive use of physics information, especially in CV tasks, as follows:

{\color{black}\begin{enumerate}

   \item \textbf{Balancing Physics and Data in Vision Models:} Vision tasks in daily scenarios heavily rely on intuitive physics, like the rules of motion and interaction. The challenge is incorporating these physics-based constraints effectively into learning frameworks, due to a lack of formalized representations. A key research area is finding the optimal balance between these constraints and data-driven approaches in computer vision models, enhancing their performance and efficiency to better manage real-world complexities.

   \item \textbf{Choice of Physics Prior:} Selecting the appropriate physics information for inclusion in PICV models requires extensive domain expertise. Whether it's using specific physical variables as network inputs \cite{wu2018seeing, du2021physics} or components of the loss function \cite{li2022using,monakhova2022dancing}, the choice of relevant variables is crucial for the success of these models.

   \item \textbf{Benchmarking and Evaluation Platforms of PICV Approaches:} PICV lacks comprehensive platforms for benchmarking and evaluation, hindering the assessment of quality and innovation in new methods. Most PICV research utilizes domain-specific datasets, complicating fair comparisons between different PICV algorithms and requiring extensive domain knowledge to understand or compare such approaches.

  \item \textbf{Scalability, Complexity, and Integration:}Developing scalable computer vision models capable of accurately predicting complex physical phenomena in dynamic environments presents significant challenges. These models necessitate a comprehensive integration of disparate physics disciplines, such as mechanics and electromagnetism, into a unified computer vision framework, ensuring coherent and effective analysis.

 \item \textbf{Handling Uncertainty and Incompleteness:} It is crucial to develop robust methods that effectively manage uncertainties and address the incompleteness of knowledge within physics-informed computer vision (PICV) models. This involves designing strategies that improve the reliability and predictability of the models under varying operational conditions.

 \item \textbf{Interpretability and Explainability:} Improving the interpretability of PI models and understanding how physical constraints interact with learned features is challenging but crucial for ensuring that models are comprehensible to both experts and laypersons.
\end{enumerate}
}

\subsection{Research gaps and future avenues:}

{\color{black} Current PICV research trends, as illustrated in Fig.~\ref{fig:PICV_prior_types_and_research_share}b, highlight the prevalent application of physics information in developing advanced forecasting and generative models, enhancing super-resolution techniques, and improving human analysis. However, areas such as classification, segmentation, and crowd analysis still underutilize physical priors. Opportunities abound for incorporating physics priors into tasks like human tracking, object detection, and video analysis.
Future PICV efforts should aim to refine inverse problem solving in high-dimensional contexts, create image generation models that accurately reflect physical realities for greater realism, and enhance predictive models for better handling of dynamic, chaotic systems. Key to this advancement is the integration of physical laws for intricate scene interpretation, such as in fluid dynamics, and the fusion of biomechanical insights with visual data for precise human motion analysis. Addressing computational efficiency, noise resilience, model generalizability, and the seamless integration of physics with machine learning is essential for the field's progression.}

}
\section{Conclusions}
This paper introduces a state-of-the-art PICV paradigm that integrates data-driven methods with insights from physics and scientific principles. We present taxonomies to classify PICV approaches by their physics information type and incorporation into the computer vision pipeline. Our review includes a variety of images from recent studies to facilitate an understanding of how physics principles are integrated into computer vision tasks. A comprehensive summary of the discussed papers is provided in Table~\ref{table: PICV characteristics}. The aim is to demystify the application of PICV methods across various domains, highlight current challenges, and inspire future research in this field.
\label{sec:conclusion}

\section{Acknowledgment} This research was partly supported by the Advance Queensland Industry Research Fellowship AQIRF024-2021RD4.

{
\bibliographystyle{ACM-Reference-Format}
\bibliography{PICV}
}

\clearpage

\end{document}